\documentclass[journal]{IEEEtran}
\usepackage{amsmath,amssymb,amsfonts}
\usepackage{algorithmic}
\usepackage{graphicx}
\usepackage{textcomp}
\usepackage{harpoon}
\usepackage{amsmath}
\usepackage{mathrsfs}
\usepackage{amssymb}
\usepackage{epstopdf}
\usepackage{subfigure}
\usepackage{bm}
\usepackage{sansmath}
\usepackage{booktabs}
\usepackage{colortbl}
\usepackage{esint}
\usepackage{stfloats}
\usepackage{lipsum}
\usepackage{multicol}
\usepackage{multirow}
\usepackage{tabularx}
\usepackage{hyperref}
\usepackage{cite}

\usepackage{algorithm}
\newtheorem{theorem}{{Theorem}}
\newtheorem{lemma}{{Lemma}}

\newtheorem{remark}{{Remark}}

\begin{document}
\title{Joint Topology and Power Optimization for Multi-UAV Collaborative Secure Communication}

\author{Bin~Qiu,~\IEEEmembership{Member,~IEEE,}
	Wenchi~Cheng,~\IEEEmembership{Senior Member,~IEEE,}
	Hongxiang~He,\\
	and~Wei~Zhang,~\IEEEmembership{Fellow,~IEEE}
\thanks{This work was supported in part by the National Natural Science Foundation of China under Grant 62341132 and Grant 62401432, in part by the National Key Research and Development Program under Grant 2024YFC3016000, and in part by the Fundamental Research Funds for the Central Universities under Grant ZYTS25266. {\emph{(Corresponding author: Wenchi Cheng.)}}}
	
	\thanks{Bin Qiu, Wenchi Cheng, and Hongxiang He are with the State Key Laboratory of Integrated Services Networks, Xidian University,
		Xian 710071, China (e-mail: qiubin@xidian.edu.cn; wccheng@xidian.edu.cn; hehongxiang@stu.xidian.edu.cn).}
	
	\thanks{Wei Zhang is with the School of Electrical Engineering and Telecommunications, University of New South Wales,
		Sydney, NSW 2052, Australia (e-mail: w.zhang@unsw.edu.au).}	
}

\maketitle

\begin{abstract}
In this paper, we investigate an unmanned aerial vehicle (UAV)-enabled secure communication scenario that a cluster of UAVs performs a virtual non-uniform linear array (NULA) to communicate with a base station (BS) in the presence of eavesdroppers (Eves). Our goal is to design the UAV topology, trajectory, and precoding to maximize the system channel capacity. To this end, we convert the original problem into equivalent two-stage problems. Specifically, we first try to maximize the channel gain by meticulously designing the UAV topology. We then study the joint optimization of the trajectory and precoding for total transmit power minimization while satisfying the constraints on providing quality of service (QoS) assurance to the BS, the leakage tolerance to Eves, the per-UAV transmit power, the initial/final locations, and the cylindrical no-fly zones. For the UAV topology design, we prove that the topology follows the Fekete-point distribution. The design of trajectory and precoding is formulated as a non-convex optimization problem which is generally intractable. Subsequently, the non-convex constraints are converted into convex terms, and a double-loop search algorithm is proposed to solve the transmit power minimization problem. Introduce random rotation offsets so as to perform a dynamic stochastic channel to enhance the security. Numerical results demonstrate the superiority of the proposed method in promoting capacity.
\end{abstract}

\begin{IEEEkeywords}
Physical layer security, non-uniform linear array, UAV communications, trajectory design, virtual antenna array.
\end{IEEEkeywords}

\IEEEpeerreviewmaketitle

\section{Introduction}
\IEEEPARstart{U}{nmanned} aerial vehicles (UAVs), commonly known as drones, have been anticipated to be crucial in both military and civilian applications due to their nature of high flexibility, low cost, and swift deployment~\cite{Survey}. Wireless transmission link is one of the key parts of UAVs. Attributing to the height of UAVs, strong line-of-sight (LoS) links are usually dominant over small-scale fading channel. In particular, the maneuverability of UAVs in a cost-effective manner facilitates information wireless services in back country and emergency scenarios. The application of UAVs will be more extensive and promote the development of different kinds of linkage industries for the foreseeable future~\cite{Comprehensive,Joint,Aware}. Notwithstanding such promising advantages and benefits, UAV-enabled communication also faces various challenges.

Security is one of the most common but serious challenges in UAV-enabled communication systems. The open transmission media and physical interception pose major threats in secure transmission of information, which makes the communication susceptible to be intercepted by eavesdroppers (Eves)~\cite{Eavesdropper}. Conventional security mechanism relied on cryptographic encryption/decryption methods, however, is inappropriate for UAV communication applications due to its high computing power requirements. Moreover, secret key distributions and managements are also difficult to apply for the flexible UAV networks. The emerging physical layer (PHY) security method is to exploit the intrinsic stochastic nature of wireless channels to strengthen the information security~\cite{PLS,MINE3}. As a great potential way to safeguard wireless communications, PHY security is an information-theoretic technique and avoids the use of computing resources. Hence, it is of interest to exploit prompt deployment and agility of UAV that provides a better role for UAV-enabled PHY security communication networks. To unlock the potential of UAV networks, trajectory design is an essential issue for PHY security provisioning. In~\cite{Robust}, robust designs of the UAV trajectory and power control were studied to maximize the average secrecy rate under worst case and outage-constrained case in a cognitive UAV communication network. Inspired by the metric of PHY security, secrecy-rate maximization criterion was adopted for the UAVs' downlink wiretap links to effectively improve the transformation performance~\cite{Game}. Another effective PHY security technique has been proposed to embed jamming noise, also called as artificial noise (AN), which is transmitted simultaneously with information signals for the purpose of degrading the channel to Eves~\cite{MINE2}. 

It is of interest to develop array techniques that provide extra security by exploiting flexibility at PHY. In array applications, the array topology is one of key techniques for system design as it has great effects on beamforming radiation pattern. A proper array topology has the capability to raise the efficiency of transceiver elements, and thus it is helpful to simplify system structure and data processing~\cite{Near_Topology}. Notably, the work in~\cite{Topologies_Spectral} analytically investigated three different array topologies, namely uniform rectangular array, uniform linear array (ULA), and uniform circular array, deployed at the base station (BS) in millimeter wave (mmWave) massive multiple-input multiple-output (mMIMO) systems, and theoretically studied the impact of array topologies on the achievable spectral efficiency. The authors in~\cite{Structures} analyzed the performance of various three-dimensional (3D) array topologies in the design of the hybrid precoder for mmWave multi-user mMIMO systems. Additionally, several tight closed-form ergodic capacity approximations for the point-to-point mmWave communication system were derived, and it is proved that they hold for an arbitrary number of paths and antennas, and arbitrary antenna array topologies~\cite{Ergodic}. As is well-known in array signal processing, more antennas equipped with the array can enhance the array’s capability in the degree of spatial freedom. On that, a novel large-scale aperiodic multibeam array topology design was presented for future 5G/6G BS to achieve more gain and much lower sidelobes~\cite{Topologies_Large}.

Caused by the size, load, and on-board energy limitations of UAVs, however, it is another challenging issue for a single UAV to realize multidimensional coverage of the mission area and high-level security. These characteristics make UAVs limited to be used in certain scenarios, e.g., a UAV neither has enough on-board energy to support effective communication with remote regions nor can it be equipped with large-scale antenna arrays to provide satisfactory performance. Upon this account, a multi-UAV collaboration network is appealing for further security enhancements. As a pioneer on UAV swarm coordination, a novel framework for deploying and operating a drone-based antenna array system whose elements are single-antenna drones for providing wireless service to a number of ground users was considered in~\cite{Drone}. The authors in~\cite{Switching} investigated a dual-UAV enabled secure communication system, where one UAV establishes reliable communication with multiple users while the other one serves as a jammer to interfere with Eves. By joint design of resource allocation, UAV trajectory, and role selection, a multi-purpose UAV scheme was proposed to provide secure communications with multiple ground users~\cite{MultiUAV}. Besides, a virtual antenna array composed of UAV swarm is an effectual way to improve the transmission performance~\cite{Distributed,Collaborative}. By leveraging the dynamical UAV swarm topology, the authors in~\cite{Stochastic} studied an analog collaborative beamforming-based PHY security approach, where a group of UAVs changes the topology in a randomly distributed manner to generate noise-like signals for distorting the received signals of the potential Eves. However, the energy efficiency of PHY security is of foremost importance task due to the inherent onboard energy limitations in UAV systems, and thus how to achieve PHY security for UAV-enable communications by employing the high mobility and flexibility with energy efficiency is of great significance.

Motivated by the aforementioned aspects, an energy efficiency multi-UAV cooperative PHY security communication scheme is proposed in this work, where a virtual non-uniform linear array (NULA) through the cooperation of multiple UAVs is formed to communicate with a multi-antenna BS. Utilizing the flexible deployment and agile mobility of UAVs, we introduce a random rotation offset "$\phi$" to construct a dynamic stochastic channel, which can provide new degrees of freedom to reduce the eavesdropping. To be specific, we aim to maximize the system channel capacity via jointly optimizing the UAV parameters---topology, trajectory, and precoding. To handle the formulated problem, the structure is first explored to gain some insights into the solutions. Concretely, it is found that the original optimization problem can be converted into equivalent two-stage problems. As such, the UAV topology is optimized by maximizing the channel gain in the first stage. In view of this, we prove that the asymptotically optimal UAV topology follows the Fekete-point distribution~\cite{Fekete}; next, in the second stage, the trajectory and transmit precoding are jointly designed to minimize the total transmit power based on the given channel matrix while meeting the quality of service (QoS), the leakage tolerance, the per-UAV power, the initial/final locations, and the no-fly zone constraints. To make a more tractable problem, after converting the non-convex constraints to convex terms and utilizing the technique of successive convex approximation (SCA), a double-loop search algorithm is proposed to find the optimal solutions of the total transmit power minimization problem. Furthermore, we extend our proposed scheme to more practical cases, such as imperfect channel state information (CSI) and high-dimensional virtual arrays.

The remainder of this paper is structured as follows. In Section II, we introduce the system model and problem statement. In Section III, we first provide some insights into the formulated problem; then, a two-stage algorithm is studied to solve the problems. Extensions to more practical cases are developed in Section IV. We discuss the simulation results in Section V, followed by the conclusions in Section VI. Some details regarding the derivations are relegated to the Appendices.

\textit{Notations}: In this paper, matrices and vectors are denoted by bold capital and lower-case letters, respectively. $( \cdot )^T$, $( \cdot )^ {-1} $, and $( \cdot )^H$ represents its transpose, inverse, and Hermitian, respectively. $\mathbb{E}\{\cdot\}$ and Tr$(\cdot)$ denote statistical expectation and trace, respectively. $\left\|  \cdot  \right\|_2$ and $\left|  \cdot  \right|$ are the Euclidean norm and absolute value, respectively. ${[ \cdot ]_{i,j}}$ stands for the $i$th row and the $j$th column element of a matrix. ${\rm{diag}}(\cdot)$ devises a diagonal matrix with the elements of its argument. We use ${\mathfrak{Re}}\{\cdot\}$ and ${\mathfrak{Im}}\{\cdot\}$ to extract the real valued part and imaginary valued part of a complex number, respectively. ${{\mathbf{I}}_N}$ indicates the identity matrix with $N \times N$. $\mathcal{O}(\cdot)$ denotes the big-O notation. We use $\mathbb{R}^{N \times M}$, $\mathbb{C}^{N \times M}$, and $\mathbb{H}^N$ to indicate the space of $N \times M$ real matrix, complex matrix, and $N \times N$ Hermitian matrix, respectively.

\section{System Model and Problem Formulation}
In this section, after introducing the system model, we present the optimization problem of interest.
\subsection{System Model}
\begin{figure}
	\centering
	\includegraphics[width=0.76\columnwidth]{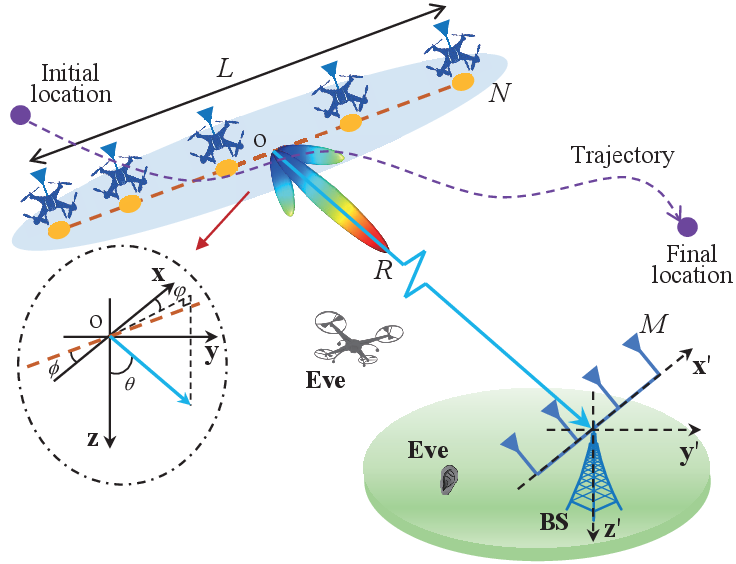}
	\caption{Illustration of the multi-UAV cooperative secure communication system.}
	\label{fig1}
\end{figure}
We consider a multi-UAV-enabled air-to-ground (A2G) wireless communication system, as illustrated in Fig.~\ref{fig1}, where a cluster of $N$ rotary-wing UAVs is deployed to communicate with a remote BS followed a collaboration protocol in the presence of $Q$ potential Eves overhearing the transmission information. More specifically, the UAVs form a transmit virtual NULA, and the BS is employed a ULA equipped with $M$ isotropic antennas. Due to the flexible deployment of UAVs and 3D free space with high altitude, it is likely that LoS links can be established for the A2G communication scenarios and thus, the link between the UAVs and the BS is assumed as a point-to-point far-field LoS channel~\cite{Modeling}. 

The virtual array topology whose elements are single-antenna UAVs refers to the geometric arrangement of multiple UAVs, which can significantly influence the beamforming capabilities, directivity, and gain. Considering this fact, we try to jointly design the topology, trajectory, and precoding to enhance the communication performance. Without loss of generality, we use a 3D Cartesian coordinate to represent the exact location of the array elements. The virtual NULA deployed by the UAVs all lies in the $x$–$y$ plane centered at the origin, which is parallel to the ground plane. Select $x$-axis as the line parallel to the received linear array. The $z$-axis is perpendicular to the $x$-$y$ plane and points to the ground. Denote by $\phi$ the angle between the transmit and received arrays. The range between the center points of the transmit and received arrays is $R$. $\varphi $ and $\theta $ represent the azimuth and the elevation angles of departure, respectively. The maximum interval between UAVs needs to be controlled within a certain range $L$ to realize information sharing. That is, assume the virtual linear array with aperture sizes $L$. For its simplicity and effectiveness, in this work, we focus on the two-dimensional (2D) trajectory design, i.e., the UAVs flight at a constant altitude, and assume that the flight altitude is chosen sufficiently high such that a LoS transmission between the UAVs and the BS holds.$\footnote{According to field measurements, for a UAV with a flight altitude of 100 meters and a cell with a radius of 600 meters, the A2G links are guaranteed to be LoS channels in rural areas~\cite{NoFly}.}$ To focus on the beamforming characteristics, we have made some assumptions to simplify the formulation, e.g., perfectly synchronized in time and frequency among the UAVs. In particular, we discretize the UAV total flight duration $T$ into $I$ time slots with small equal-duration $T/I$, and each time slot index is given by $i \in {\cal I}\buildrel \Delta \over = \{1,2,...,I\}$. Hereby, the MIMO transmission in any time slot is modeled as$\footnote{Since the channel and received signal are repeated at each time slots, for notational simplicity, we drop time slot index $i \in {\cal I}$ when there is no ambiguity.}$ 
\begin{align}
{\bf{y}} = {\bf{Hs}} + {\bf{n}},
	\label{eq1}\end{align}
where ${\bf{y}} \in \mathbb{C}^{M \times 1}$ denotes the received signal vector; ${\bf{n}}\in \mathbb{C}^{M \times 1}$ is the complex additive white Gaussian noise (AWGN) with zero mean and variance $\sigma^2$, satisfying ${\bf{n}}\sim{\mathcal{CN}}({{\boldsymbol{0}}},\sigma^2{\bf{I}}_M)$; ${\bf{s}}={\bf{Wx}} = \sum\nolimits_{k \in {\cal K}} {{{\bf{w}}_k}{x_k}}\in \mathbb{C}^{N \times 1}$ denotes the transmit signal vector with ${\bf{W}}\buildrel \Delta \over = [{{\bf{w}}_1},{{\bf{w}}_2},...,{{\bf{w}}_K}]\in \mathbb{C}^{N \times K}$ being the precoding matrix, which processes $K$ $(K \le M)$ transmit symbol streams ${\bf{x}} = [{x_1},{x_2},...,{x_K}]^T \in \mathbb{C}^{K \times 1}$ to match the antenna array, satisfying $\mathbb{E}\{{{\bf xx}^H}\} = {\bf I}_K$,$\footnote{To represent a more general scene, we focus on spatial multiplexing, and the results can also be applied to the spatial diversity by modifying the transmit signal with ${\bf{s}} = {\bf{w}}x.$}$ with ${\cal K} \buildrel \Delta \over = \{ 1,2,...,K\}$; ${\bf{H}} = \{ {h_{m,n}}\}\in \mathbb{C}^{M \times N}$, $\forall m \in {\cal M}$, $\forall n \in {\cal N}$, denotes the channel matrix with ${\cal M}\buildrel \Delta \over = \{ 1,2,...,M\}$ and ${\cal N} \buildrel \Delta \over= \{ 1,2,...,N\}$. According to the ray tracing principle, the channel coefficient between each transmit-receive antenna pair is a deterministic value of the range between them, which is given by~\cite{High_Rank}
\begin{align}
{h_{m,n}} \approx \rho (R){e^{ - j2\pi {f_c}(t - \frac{{\tau_{m,n}}}{c})}}, \forall n\in {\cal N}, \forall m\in {\cal M},
	\label{eq2}\end{align}
where ${\tau_{m,n}}$ is the wave propagation range from $n$th transmit UAV to $m$th received element; $\rho (R) = \frac{c}{{4\pi {f_c}R}}$ is the signal attenuation factor caused by loss of free-space path for radio propagation related to $R$. Considering the assumption of far-field transmission, we can ignore the difference in attenuation among array elements, i.e., $\rho(\tau_{m,n})  \approx \rho (R)$. Denote by $\{\eta_n\}_{n \in {\cal N}} \in [-1,1]$ the normalized spacing on the transmit virtual NULA. In general, the UAVs are assumed to be equipped with Global Positioning System (GPS) to get information regarding own positions. Thus, the $n$th UAV element position relative to its center is $L\eta_n/2$, $\forall n\in {\cal N}$. Therefore, the specific coordinate of the $n$th transmit UAV, $\forall n\in {\cal N}$, denoted by ${\bf p}_{U,n}=({x_{U,n}},y_{U,n},z_{U,n})$, is given by 
\begin{align}
\left\{ \begin{array}{l}
	{x_{U,n}} = \frac{{L\eta_n \cos \phi }}{2},\\
	{y_{U,n}} = \frac{{L\eta_n \sin \phi }}{2},\\
	{z_{U,n}} = 0.
\end{array} \right.
	\label{eq3}\end{align}
The instantaneous coordinate of the $m$th received antenna at the BS, $\forall m\in {\cal M}$, denoted by ${\bf p}_{B,m}=(x_{B,m},y_{B,m},z_{B,m})$, is given by 
\begin{align}
\left\{ \begin{array}{l}
	{x_{B,m}} = \frac{{2m - 1 - M}}{2}d + R\sin \theta \cos \varphi,\\
	{y_{B,m}} = R\sin \theta \sin \varphi,\\
	{z_{B,m}} = R\cos \theta,
\end{array} \right.
	\label{eq4}\end{align}
where $d$ represents the inter-element spacing of the received ULA. Therefore, the specific wave propagation range ${\tau_{m,n}}$ can be expressed as
\begin{multline}
	{\tau _{m,n}} \!=\! \sqrt {{{({x_{B,m}} \!-\! {x_{U,n}})}^2} \!+ \!{{({y_{B,m}} \!-\! {y_{U,n}})}^2} \!+\! {{({z_{B,m}}\! -\! {z_{U,n}})}^2}} \\
	\kern 22pt   \approx  R  \!+ \! \frac{{{{(2m \! -  \!1 \! - \! M)}^2}{d^2}}}{{8R}} \! + \! \frac{{(2m \! - \! 1 - \! M)d\sin \theta \cos \varphi }}{2} \\
	\kern 40pt - \frac{{2m - 1 - M}}{{4R}}dL{\eta _n}\cos \phi - \frac{{L{\eta _n}\sin \theta \cos \varphi \cos \phi }}{2} \\
	- \frac{{L{\eta _n}\sin \theta \sin \varphi \sin \phi }}{2} + \frac{{{{(L{\eta _n})}^2}}}{{8R}},\kern 47pt 
	\label{eq5}\end{multline}
where the approximation is proper by processing a Maclaurin series expansion on the first order~\cite{Mathematics}, i.e. ${(1 + x)^{1/2}} \approx 1 + x/2$.

\begin{remark} \label{remark01}
Suppose that: (i) the spacing among UAVs is sufficiently separated such that mutual coupling effects can be negligible; (ii) each UAV is served as an ideal isotropic single antenna; (iii) UAVs share the rotation offset and the location information, whereas any information of Eves is not available; (iv) the BS and Eves are located in the different places, thus PHY security technique can guarantee high secrecy rate.
\end{remark}
\subsection{Problem Formulation}
We consider the case that Eves try to passively eavesdrop the confidential messages. That is, they keep the radio silent so as not to cause any interference. Upon this account, any information of wiretap channels is not available. As is well known, information theory was invented by Claude Shannon to characterize the limits of reliable communication~\cite{Capacity}. Our target is to achieve reliable communication as well as PHY security. To this end, we employ the multi-UAV collaborative strategy to maximize the system channel capacity by jointly designing the topology, trajectory, and precoding, i.e.,
\begin{align}
\textbf{P1:}\kern 4pt\mathop {\max }\limits_{\{ {\boldsymbol{\eta }},{\bf{W}},{\bf{p}}_U'\} }\quad C = \sum\limits_{k \in{\cal K}} {{{\log }_2}} \left( {1 + \frac{\gamma }{N}{\lambda _k}} \right),
	\label{eq6}\end{align}	
where ${\lambda _k}$ is the $k$th largest eigenvalue of the channel gain matrix ${\bf{\Pi}} \in \mathbb{H}^M$ defined as ${\bf{\Pi }} \buildrel \Delta \over = {\bf{H}}{{\bf{H}}^H}$;$\footnote{Here we have assumed $N \ge M$. Otherwise, the channel gain matrix is defined as ${\bf{\Pi }} \buildrel \Delta \over = {\bf{H}}^H{{\bf{H}}}$.}$ $\boldsymbol{\eta }\buildrel \Delta \over=[\eta_1,\eta_2,...,\eta_N]^T  \in \mathbb{R}^{N \times 1}$ is the normalized spacing vector; and ${\gamma}$ denotes average received signal-to-noise ratio (SNR). For simplicity, all the average received SNRs of each received symbol streams are assumed to be identical, i.e., ${\gamma}={\gamma_k}, \forall k$.
\section{A Two-stage algorithm for Optimization Problem}
To achieve PHY security, it is necessary to ensure that messages can be reliably transmitted to the BS while avoiding information interception. Let us give some insights of \textbf{P1}. According to the insights, the original problem is then converted into equivalent two-stage problems. 

\subsection{Some Insights to the Problem}
For the monotonicity of logarithmic function, we further simplify the objective function in \textbf{P1} without affecting optimality, i.e.,
\begin{eqnarray}
	\begin{aligned}[b]
	C &\to \sum\limits_{k \in {\cal K}} {{{\log }_2}} \left( {\frac{\gamma }{N}{\lambda _k}} \right)\\&
	= K{\log _2}\left( {\frac{\gamma }{N}} \right)\; + {\log _2}\left( {\prod\limits_{k \in {\cal K}} {{\lambda _k}} } \right).
\end{aligned}
\label{eq7}\end{eqnarray}
Based on the above transformation, it can be found that the first term is related to the received SNR, and the second term is determined by the eigenvalue product of channel gain matrix. As a consequence, we convert the optimal problem into equivalent two-stage problems. In the first stage, we aim to maximize the eigenvalue product of channel gain matrix via designing the UAV topology; next, based on the channel matrix formed by the given topology, we try to minimize the total transmit power subject to meeting the QoS requirements by jointly optimizing the trajectory and precoding in the second stage.

The pure LoS MIMO channel matrix becomes high rank by meticulous design of the topology to provide the capacity gain~\cite{LoS_MIMO,High_Rank}. In view of this, the instantaneous eigenvalue product maximization problem related to the channel, in the first stage, is then given as  
\begin{align}
\textbf{P2:}\kern 4pt \mathop {\max }\limits_{ \boldsymbol{\eta }} \prod\limits_{k \in {\cal K}} \kern 2pt{{\lambda _k}}.
	\label{eq8}\end{align}
Let us first consider an equivalent channel. Inserting \eqref{eq5} into  \eqref{eq2}, the channel matrix can be rewritten as
\begin{align}
	{\bf{H}} ={e^{ - j2\pi {f_c}(t - \frac{R}{c})}} \rho (R){{\bf{G}}_B}{\bf{\tilde H}}{{\bf{G}}_U},
	\label{eq9}\end{align}
where ${{\bf{{G}}}_B} = {\rm{diag}}({g_{B,1}},{g_{B,2}},...,{g_{B,M}})$ with ${g_{B,m}} = {e^{ - j\frac{{2\pi {f_c}}}{c}[\frac{{{{(2m - 1 - M)}^2}{d^2}}}{{8R}} + \frac{{(2m - 1 - M)d\sin \theta \cos \varphi }}{2}]}}$, $\forall m\in {\cal M}$; ${{\bf{G}}_U} = {\rm{diag}}({g_{U,1}},{g_{U,2}},...,{g_{U,N}})$ with ${g_{U,n}} = {e^{ - j\frac{{2\pi {f_c}}}{c}[ {\frac{{{{(L{\eta _n})}^2}}}{{8R}} - \frac{{L{\eta _n}\sin \theta \cos \varphi \cos \phi }}{2} - \frac{{L{\eta _n}\sin \theta \sin \varphi \sin \phi }}{2}}]}}$, $\forall n\in {\cal N}$; ${\bf{\tilde H}} = \{ {\tilde h_{m,n}}\}\in \mathbb{C}^{M \times N}$ denotes the asymptotic channel matrix whose element can be expressed as
\begin{eqnarray}
	\begin{aligned}[b]
		{\tilde h_{m,n}} = {e^{j\nu \frac{{(2m - 1 - M)}}{M}{\eta _n}\cos \phi }}, \forall n\in {\cal N}, \forall m\in {\cal M},
	\end{aligned}
	\label{eq10}\end{eqnarray}
with $\nu  = \frac{\pi f_c}{c}\frac{dML}{2R}$. Note that the singular values of channel matrix ${\bf{H}}$ are identical to those of asymptotic channel matrix ${\bf{\tilde H}}$ since both ${{\bf{G}}_U}$ and  ${{\bf{G}}_B}$ in \eqref{eq9} are unitary by definition. That is, replacing ${\bf{\Pi}}$ by ${\bf{\tilde \Pi}}\buildrel \Delta \over = {\bf{\tilde H}}{{\bf{\tilde H}}^H}$ yields a simpler way to calculate the eigenvalues.

There are two ways to reduce the probability of eavesdropping: (i) randomly scrambling the received signal of Eves (dynamic security); (ii) reducing the SNR of the received signal of Eves (static security). Accordingly, we design an energy-efficient secure transmission strategy suitable for UAVs based on the two aforementioned approaches.

\begin{figure}
	\centering
	\includegraphics[width=0.75\columnwidth]{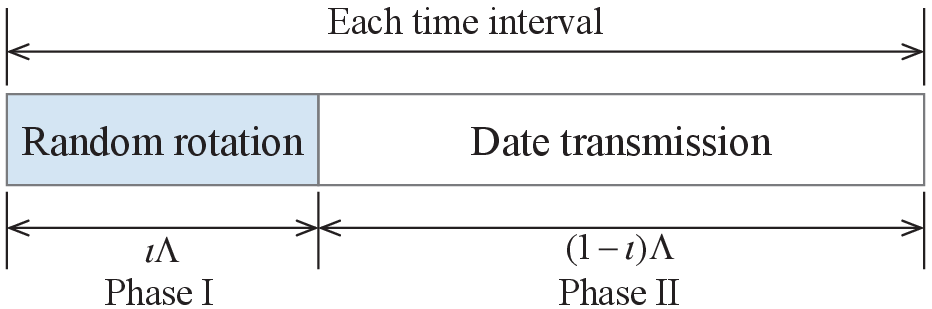}
	\caption{Two phases for the multi-UAV cooperative secure communication.} 
	\label{fig2}
\end{figure}

The AN-aided PHY security is a feasible way to degrade the channel of Eves. However, the insert of AN inevitably consumes a proportion of the transmit power and consequently makes power less effective. Because UAVs are typically battery-powered, UAV-enabled communication systems are inherently limited by their onboard energy. Besides, UAVs with high mobility and flexibility can provide new degrees of freedom to form dynamic transmission channels. Driven by the UAVs' characteristics, we introduce a random rotation offset, i.e., the angle $\phi$, to form dynamic stochastic channels to randomly scramble the received signals of Eves such that enhance the security. The multi-UAV cooperation strategy follows a protocol with two phases, as shown in Fig.~\ref{fig2}. Within each predefined time interval period of length ${\Lambda}$, the UAVs randomly change the rotation offsets and transmit via cooperation. Specifically, in the random positioning phase I for ${\iota \Lambda}$ with ${0 < \iota  < 1}$, the UAVs first randomly alter the rotation offsets under center hover; In the subsequent data transmission phase II for ${(1-\iota) \Lambda}$, the UAVs fly on a trajectory and adjust precoding to get constructive synthesis beamforming for the BS. The dynamic stochastic channel conduces a noise-like signal for the Eves.

Denote by ${\bf{p}}'_U$ and ${\bf{p}}'_B$ the center coordinates of the transmit virtual NULA and the received ULA, respectively. In particular, the maximum UAV flight speed is bounded below $V_{\rm max}$, and the initial and final center coordinates of the UAVs are determined at ${\bf d}_{I}$ and ${\bf d}_{F}$, respectively. In practical application, there exist some no-fly zones that prohibit UAV flights, such as airports, jurisdictions, military districts, etc~\cite{NoFly}. As a result, for security reasons, no-fly zones are commonly imposed on UAVs. We assume the existence of a cylindrical no-fly zone with radius $\Upsilon$ and center coordinate ${\boldsymbol \chi}$. We note that each UAV has its own power limitations. Thus, one may need to limit the per UAV peak power to operate within the linear region of the power amplifier~\cite{Cognitive}. Let $P_{\rm max}$ denote the maximum transmit power for each UAV. Since the displacement of the UAV in each time slot is relatively small, we assume that the channels remain unchanged during one time slot. Hence, the UAV trajectory and the beamforming policy in time slot are designed based on the channels at the end of time slot. This procedure is repeated for time slots, and the whole UAV trajectory is obtained by combining the respective time slot trajectories. On the other hand, considering an energy-efficient design is of utmost importance for UAV-enable communication systems, we jointly design the UAV trajectory and precoding to minimize the total transmit power, which not only saves power consumption, but reduces the possibility of information leakage. The total transmit power over all time slots is defined as $\Gamma \buildrel \Delta \over = \sum\nolimits_{i \in {\cal I}} {{\rm{Tr}}({\bf{W}}[i]{{\bf{W}}^H}[i])}$. In the second stage, with given channels obtained in the first stage, the total transmit power minimization problem is formulated as
\begin{align}
	\textbf{P3:}\kern 4pt &\mathop {\min }\limits_{\{ {\bf{W}},{{\bf{p}}^\prime_U}\} }\kern 2pt \Gamma \tag{11a} \label{eq11a}\\&
	{\rm{s.t.}} \kern 4pt  {\rm{SNR}}_{B,k} \ge \gamma ,\forall k \in {\cal K}, \tag{11b} \label{eq11b}\\&
\kern 17pt	{\rm{Pr}}\left(\mathop {{\rm{max}}}\limits_{q \in {\cal Q}}\kern 2pt  {\rm{ SNR}}_{E,q} \le \xi\right ) \ge \kappa, \tag{11c} \label{eq11c}\\&
\kern 17pt	{\left[{\bf{W}}{{\bf{W}}^H}\right]_{n,n}} \le {P_{\rm max}},\forall n \in {\cal N}, \tag{11d} \label{eq11d}\\&
\kern 17pt	\|{{\bf{p}}^\prime_U}-{{\bf{p}}^\prime_U}[i-1] \|_2^2 \le \delta V_{\rm max} , \tag{11e} \label{eq11e}\\&
\kern 17pt	\|{{\bf{p}}^\prime_U}-{\boldsymbol \chi } \|_2^2 \le \Upsilon^2, \tag{11f} \label{eq11f}\\&
\kern 17pt	{{\bf{p}}^\prime_U}[0] = {\bf d}_{I},{{\bf{p}}^\prime_U}[T] = {\bf d}_{F}, \tag{11g} \label{eq11g}
\end{align}
where ${\rm{SNR}}_{B,k}=\frac{{{{\left| {{{\bf{h}}_k^H}{{\bf{w}}_k}} \right|}^2}}}{{\sum\nolimits_{i \in {\cal K}\backslash\{k\}} {{{\left| {{{\bf{h}}_k^H}{{\bf{w}}_i}} \right|}^2} + {\sigma ^2}} }}$ is the received SNR of the $k$th symbol stream with ${{\bf{h}}_k} \in {[{\bf{h}}_1,{\bf{h}}_2,...,{\bf{h}}_M]} = {\bf{H}}^H$; $\gamma$ denotes the prescribed minimum SNR required for QoS; ${\rm{ SNR}}_{E,q}={{\sum\nolimits_{k \in {\cal K}} {{{\left| {{{\bf{h}}_{E,q}^H}{{\bf{w}}_k}} \right|}^2}} }}/{{\sigma _E^2}}$ is the upper bound of the received SNR for the $q$th Eve \cite{Duplex}, with ${{\bf{h}}_{E,q}}$ being the associated $q$th Eve channel and ${\sigma _E^2}$ being the noise power at Eves; $\xi$ represents the maximum leakage SNR tolerance; $\kappa$ represents a probabilistic parameter, which means that the maximum received SNR for all Eves is limited to less than the maximum tolerable leakage SNR $\xi$ within at least probability $\kappa$. Our design is based on providing QoS assurance to each of the data streams. Since the received SNR essentially determines the maximum achievable data rate and the probability of error, it is an effective measure of QoS. The constraint in~\eqref{eq11b} is to protect the UAVs toward the BS links. The constraint in~\eqref{eq11c} picks out the minimum outage requirements for all Eves.$\footnote{Besides, although the number of Eves is not available, $Q$ indicates the system with the capability of processing maximum number of Eves.}$ The constraint in \eqref{eq11d} is imposed such that the power radiated by per UAV puts bounds to $P_{\rm max}$. The constraints in~\eqref{eq11e}-~\eqref{eq11g} together restrict the trajectory design.

\subsection{Virtual NULA Topology Design} 
Now, our main task in this section is to derive an optimization algorithm to solve \textbf{P2}. Let us consider the case that the communication distance is sufficiently large compared to the product of the antenna aperture sizes of the transmit and receive array, i.e., $\nu \to 0$, which facilitates our optimization problem formulation. Then, we transform the eigenvalue of channel gain matrix using the following theorem. 
 
\begin{theorem} \label{theorem01}
When $\nu \to 0$, the $k$th largest eigenvalue of the channel gain matrix, $\forall k \in {\cal K}$, can be asymptotically expressed as
\begin{align}
		\setcounter{equation}{11}
	{\lambda _k} \approx {\left[ {\frac{{{r_{B,k}}{r_{U,k}}}}{{(k - 1)!}}}\right]^2}{(\nu \cos \phi )^{2(k - 1)}},
	\label{eq12}\end{align}
where ${r_{B,k}}$ and ${r_{U,k}}$ denote the $k$th diagonal entries of the upper triangular matrices given in~\eqref{eq56}.	
\end{theorem}

\begin{IEEEproof}
	Please refer to Appendix A.
\end{IEEEproof}	
	
Inserting~\eqref{eq12} into~\eqref{eq8}, the objective function is given by
\begin{eqnarray}
	\begin{aligned}[b]
	  \prod\limits_{k \in {\cal K}} \kern 2pt{{\lambda _k}}& \approx \prod\limits_{k \in {\cal K}} \kern 2pt{\left[{\frac{{{r_{B,k}}{r_{U,k}}}}{{(k - 1)!}}}\right]^2}{(\nu \cos \phi )^{2(k - 1)}}\\&
	  =\prod\limits_{k \in {\cal K}}r_{U,k}^2\prod\limits_{k \in {\cal K}}{\left[{{{\frac{{{r_{B,k}}(\nu \cos \phi )}}{{(k - 1)!}}}^{k - 1}}}\right]^2}.
	\end{aligned}
\label{eq13}\end{eqnarray}
Then, \textbf{P2} can be equivalently expressed as
\begin{align}
	\textbf{P2.1:}\kern 4pt \mathop {\max }\limits_{{\boldsymbol{\eta }}} \prod\limits_{k \in {\cal K}} \kern 2pt{r_{U,k}^2}.
	\label{eq14}\end{align}
To find a closed-form relationship between ${\{r_{U,k}\}_{k \in {\cal K}}}$ and ${{\{ {\eta _n}\} }_{n \in {\cal N}}}$, we introduce the following Theorem.

\begin{theorem} \label{theorem02}
The diagonal entries of upper-triangular matrix can be further expressed as
\begin{align}
	\left\{ \begin{array}{l}
	{r_{U,1}^{2}} =  N, \\
	{r_{U,k}^{2}} = {\frac{{\sum\limits_{{\cal S}_k} {\prod\limits_{\{a < b\}\in {{\cal S}_k} } {({\eta _b} - {\eta _a})}^2} }}{{\sum\limits_{{\cal S}_{k - 1}} {\prod\limits_{\{a < b\}\in  {{\cal S}_{k-1}}} {({\eta _b} - {\eta _a})}^2}} }}
\end{array} \right.,\forall k > 1,
	\label{eq15}\end{align}
where ${{\cal S}_k}$ is the set of any $k$ combinations, i.e., a subset of $N$ with size $k$, and there are $\binom{N}{k}$ of them. 
\end{theorem}

\begin{IEEEproof}
	Please refer to Appendix B.
\end{IEEEproof}	
According to Theorem ~\ref{theorem02}, the objective function in \textbf{P2.1} can be written as
\begin{eqnarray}
	\begin{aligned}[b]
	\prod\limits_{k \in {\cal K}} {r_{U,k}^2} & = N\prod\limits_{k \in {\cal K}\atop k \ne 1} {\frac{{\sum\limits_{{{\cal S}_k}} {\prod\limits_{\{a < b\}\in {{\cal S}_k}} {({\eta _b} - {\eta _a})}^2}}}{{\sum\limits_{{\cal S}_{k - 1}} {\prod\limits_{\{a < b\}\in {{\cal S}_{k-1}}} {({\eta _b} - {\eta _a})}^2}}}} \\&
	= \sum\limits_{{\cal S}_K} {\prod\limits_{\{a < b\} \in {{\cal S}_K}} {{({\eta _b} - {\eta _a})}^2}}.
	\end{aligned}
\label{eq16}\end{eqnarray}
Inserting~\eqref{eq16} into~\eqref{eq14}, \textbf{P2.1} can be recast as
\begin{align}
	\textbf{P2.2:}\kern 4pt \mathop {\max }\limits_{{\boldsymbol{\eta}}} {\mathscr{F}_{{\cal S}_K} ({\boldsymbol{\eta}})},
	\label{eq17}\end{align}
where
\begin{align}
	{\mathscr{F}}_{{\cal S}_K}({\boldsymbol{\eta}})= \sum\limits_{{\cal S}_K} {\prod\limits_{\{a < b\} \in {\cal S}_K} {{{({\eta _b} - {\eta _a})}^2}}}. 
	\label{eq18}\end{align}	
It is important to note that the number of UAVs should be greater than or equal to that of the data streams. Then, we can present this in two cases: $K = N$ and $K < N$.

When $K = N$, the objective function in \textbf{P2.2} reduces to
\begin{align}
	{\mathscr{F}}_{{\cal S}_K}({\boldsymbol{\eta}})={\prod\limits_{\{a < b\}\in{{\cal S}_K}} {{{({\eta _b} - {\eta _a})}^2}}}. 
	\label{eq19}\end{align}
It can be found that ${\mathscr{F}}_{{\cal S}_K}({\boldsymbol{\eta}})$ is the squared determinant of the Vandermonde matrix with respect to UAV topology ${\boldsymbol{\eta}}$. The optimization problem is known as the corresponding Vandermonde determinant maximization problem~\cite{Variables}. Such a problem was first introduced in~\cite{Koeffizienten}. The optimal solutions of $\{ {\eta _n}\}_{\forall n \in  {\cal N}}$ are referred to as Fekete-point distribution~\cite{Fekete}.

Further, we introduce the following Lemma to explain the case of $K < N$

\begin{lemma} \label{lemma01}
Divide $N$ UAVs into $K$ groups, and the $k$th group satisfies
\begin{align}
\{{\eta _n}\} = \{{\beta  _k}\},\quad {\rm{if}} \ k - 1 < nK/N \le k,
	\label{eq20}\end{align}
where $\{{\beta _k}\}_{k \in  {\cal K}}$ denote the $K$ Fekete points.
\end{lemma}

\begin{IEEEproof}
	Please refer to Appendix C.
\end{IEEEproof}	

\begin{remark} \label{remark02}
Above procedure reveals an interesting fact that the eigenvalue product of the asymptotic channel matrix is only related to the UAV topology. In other words, the maximum eigenvalue product is independent of time and space. This fact facilitates a stand-alone design of UAV topology. Meanwhile, it is reasonable to deploy the received ULA as Fekete-point distribution to further increase the capacity.
\end{remark}

\subsection{Joint Trajectory and Precoding Design} 
It can be observed that constraint~\eqref{eq11c} is the probabilistic constraint involving tightly coupled optimization variables. Considering a great many reflected and scattered routes between UAVs and Eves, the Eve channels can be modeled as independent and identically distributed (i.i.d.) Rayleigh fading channels based on the central limit theorem~\cite{Fundamentals}. We give the following lemma to transform the probabilistic constraint into a linear matrix inequality.	

\begin{lemma} \label{lemma02}
The probabilistic constraint~\eqref{eq11c} can be transformed as
\begin{align}
	&\sum\limits_{k \in \mathcal{K}} {{\bf{W}}_k}  \preceq {{\bf{I}}_N} \Phi _N^{ - 1}\left( {1 - {\kappa ^{1/Q}}} \right){\xi}\sigma _E^2\label{eq21}\\&
	\Rightarrow \Pr \left( {\mathop {\max }\limits_{q \in \mathcal{Q}}\kern 2pt {{\rm SNR} _{E,q}} \le {\xi}} \right) \ge \kappa,
	\label{eq22}\end{align}
where ${{\bf{W}}_k} \buildrel \Delta \over = {{\bf{w}}_k}{{\bf{w}}_k^H}$, and $\Phi _N^{ - 1}\left(  \cdot  \right)$ denotes the inverse cumulative distribution function (c.d.f.) of an inverse central chi-square random variable with $2N$ degrees of freedom.
\end{lemma}

\begin{IEEEproof}
	The proof is given in Appendix D.
\end{IEEEproof}

\begin{remark} \label{remark03}
We would like to emphasize that the transformation in~\eqref{eq21} can be applied to any continuous channels by replacing an inverse c.d.f. with the corresponding distribution.
\end{remark}

\begin{remark} \label{remark04}
It is noteworthy that the transformation holds, but not vice versa, as the inequality in~\eqref{eq74}. Consequently,~\eqref{eq21} is a tightness of~\eqref{eq22}. That is, replacing constraint~\eqref{eq22} with~\eqref{eq21} generates a small feasible solution set.
\end{remark}

One can easily verify that constraint~\eqref{eq21} is tractable in a sense that: i) a feasible solution satisfying~\eqref{eq21} is also feasible for constraint~\eqref{eq22}; and ii) the converted constraint in~\eqref{eq21} is a convex linear matrix inequality constraint.

Note that the problem in \textbf{P3} is still intractable due to the non-convexity of no-fly zone constraint~\eqref{eq11f}. Therefore, we leverage the SCA technique to derive its convex approximation as~\cite{MultiUAV}
\begin{align}
\| {{{\bf{p}}^{\prime(s)}_U} - {\boldsymbol \chi}} \|_2^2+2({{\bf{p}}^{\prime(s)}_U}-{\boldsymbol \chi})^T({{\bf{p}}^{\prime}_U}-{{\bf{p}}^{\prime(s)}_U})\ge {\Upsilon ^2}.
	\label{eq23}\end{align}
Thus, replacing constraints~\eqref{eq11c} and~\eqref{eq11f} by~\eqref{eq21} and~\eqref{eq23}, the optimization problem can be expressed as
\begin{align}
	\textbf{P3.1:}\kern 4pt &\mathop {\min }\limits_{\{ {\bf{W}},{{\bf{p}}^\prime_U}\} }\kern 2pt {\Gamma} \tag{24a} \label{eq24a}\\&
	{\rm{s.t.}} \kern 4pt {\bf{W}}{{\bf{W}}^H}  \preceq {{\bf{I}}_N} \Phi _N^{ - 1}\left( {1 - {\kappa ^{1/Q}}} \right){\xi}\sigma _E^2, \tag{24b} \label{eq24b}\\&
	\kern 17pt	{{\rm{Tr}}\left( {{{\bf{E}}^{(n)}}{\bf{W}}{{\bf{W}}^H}} \right)}  \le {P_{\rm max}},\forall n \in {\cal N}, \tag{24c} \label{eq24c}\\&
	\kern 17pt	\| {{{\bf{p}}^{\prime(s)}_U} - {\boldsymbol \chi}} \|_2^2+2({{\bf{p}}^{\prime(s)}_U}-{\boldsymbol \chi})^T({{\bf{p}}^{\prime}_U}-{{\bf{p}}^{\prime(s)}_U})\ge {\Upsilon ^2}, \tag{24d} \label{eq24d}\\&
	\kern 17pt  \eqref{eq11b},  \eqref{eq11e}, \eqref{eq11g}, \tag{24e} \label{eq24e}
\end{align}
where ${{\bf{E}}^{(n)}}  = {\bf{e}}_n{\bf{e}}_n^H$ with ${\bf{e}}_n \in {\mathbb{R}^{ N  \times 1}}$ satisfying ${\bf{e}}_n=[\boldsymbol{0}_{(n-1) \times 1}^T,1,\boldsymbol{0}_{(N-n)\times 1}^T]^T$.
\begin{algorithm}[t]
	\label{Code:1}
	\caption{Double-Loop Search Algorithm for Solving \textbf{P3.1}}
	{\bf Initialization:} Set $\{{\bf{p}}^{\prime(0)}_U\}$, $i:=0$, $s:=0$, $T$, and $\epsilon_{out}>0$;
	\begin{algorithmic}[1]
		\STATE \textbf{repeat}\{Outer iteration\};
		\STATE  ~~$s=s+1$;
		\STATE  ~~\textbf{repeat}\{Inner iteration\}
		\STATE ~~   \kern 4pt $i=i+1$;
	    \STATE ~~ 	\kern 4pt Solve \textbf{P3.1} via CVX under ${{\bf{p}}^{\prime(s-1)}_U}$;
		\STATE ~~ 	\kern 4pt Calculate channel matrix ${\bf H}[i]$ by \eqref{eq5} based on the \\ ~~ 	\kern 4pt current UAV location ${{\bf{p}}^{\prime(s-1)}_U}[i]$;
		\STATE ~~ 	\kern 4pt Determine ${{\bf{W}}}[i]$ based on channel matrix ${\bf H}[i]$;
		\STATE ~~ 	\kern 4pt Obtain UAV trajectory ${\bf{p}}^{\prime}_U[i]$;
		\STATE ~~\textbf{until}  $i=I$;
		\STATE ~~Update $\{{\bf{p}}^{\prime(s)}_U[i]\}_{i\in {\cal I}}$ = $\{{\bf{p}}^{\prime}_U[i]\}_{i\in {\cal I}}$;
		\STATE \textbf{until} $\| {{\bf{p}}^{\prime(s)}_U}[i] - {{{\bf{p}}^{\prime(s-1)}_U}}[i] \|_2^2<\epsilon_{out},\forall i$;
		\STATE \textbf{Output:} Get the finally optimal solution $(\{{{\bf{W}}^\star}\}, \{{{\bf{p}}^{\prime\star}_U}\})$.
	\end{algorithmic}
\end{algorithm}

The \textbf{P3.1} now is expressed in a standard semidefinite programming problem (SDP) form, which can be solved via a double-loop search as in Algorithm 1. Specifically, in the inner loop, using the convex solver such as CVX~\cite{Convex}, the trajectory and precoding are conducted over all time slots. For each trajectory found in the inner loop, check whether termination condition is satisfied in the outer loop. 
\subsection{Computational Complexity Analysis}
In this subsection, the complexity of the proposed scheme is analyzed. Specifically, the proposed scheme consists of three-part designs: topology, trajectory, and precoding. According to the Remark~\ref{remark02}, we know that the topology design is time and space independent. The complexity of topology design is $\mathcal{O}(1)$. The main complexity of the joint trajectory and precoding design comes from solving {\bf P3.1} via double-loop search algorithm. In the inner loop, using the convex solver based on the interior-point method, the individual complexity can be calculated by $\mathcal{O}\left((N^2I+2I)^{3.5} \log (1/\epsilon_{in} )\right)$, given the solution accuracy of $\epsilon_{in}>0$~\cite{Interior}. Accounting for the SCA, the complexity in the outer loop is of the order $\mathcal{O}\left(\log (1/\epsilon_{out} )\right)$~\cite{SCA}. Besides, updating
$\{{{\bf{w}}_k}\}_{k \in \cal K}$ by eigenvalue decomposition needs the complexity of $\mathcal{O}\left(KN^3\right)$. Hence, the complexity of the overall procedure is $\mathcal{O}\left((N^2I+2I)^{3.5} \log (1/\epsilon_{in} )\log (1/\epsilon_{out} )+\left(KN^3\right)\right)$.
	
\section{Some Extensions}
As discussed previously, the design center has focused on the perfect CSI and one-dimensional virtual NULA to simplify the model. More practical cases, such as imperfect CSI and high-dimensional virtual NULA, are presented in this section.

\subsection{Robust Design for Imperfect CSI}
The feasibility of the problem depends on the high-accuracy positioning technique and the robust design methods. To meet the high accuracy requirements for position, several advanced technologies and methodologies can be incorporated, such as high-precision GPS, lidar-based positioning, ultra-wideband positioning, and AI-based position control. This aspect is beyond the scope of this paper and thus, it will not be discussed further. Then, we focus on the robust design methods. In practical UAV-enable communication systems, UAV-mounted transmitters flying in the sky commonly encounter strong wind which leads to non-negligible body jittering. In addition, the perfect knowledge of the BS location cannot be acquired due to the limited accuracy of positioning modules. All of these factors can degrade transmission performance.  We consider the imperfect CSI condition caused by estimation location errors and UAVs jittering. On account of this, based on the bounded location error model, the channel is given by ${{\bf{h}}_{m}} = {{\bf{\hat h}}_{m}} + \Delta {{\bf{h}}_{m}},  \forall m \in \mathcal{M}$, where ${{\bf{\hat h}}_{m}}$ denotes the estimation channel, and $\Delta {{\bf{h}}_{m}}$ denotes the CSI uncertainty. It is assumed that the uncertainty is independent and equal variance~\cite{MINE1}. A set of all uncertainty, denoted by ${\Omega _m}$, is given by 
\begin{align}
		\setcounter{equation}{24}
	{\Omega _m}  = \left\{ {\Delta {{\bf{h}}_{m}}{\in \mathbb{C}^{N \times 1}}: \Delta {\bf{h}}_{m}^H\Delta {{\bf{h}}_{m}} \le \varepsilon_m^2} \right\},\forall m \in \mathcal{M},
	\label{eq25}\end{align}
where $\varepsilon_m$ is the radius of the uncertainty. The constraint~\eqref{eq11b} imposed on such a given CSI uncertainty is altered to
\begin{align}
\mathop {{\rm{min}}}\limits_{\Delta {{\bf{h}}_k} \in {\Omega _k}} {\rm{SNR}}_{B,k} \ge \gamma ,\forall k \in {\cal K}.
	\label{eq26}\end{align}
	
To facilitate the solutions, we transform constraint~\eqref{eq26} into a linear matrix inequality using the following lemma.

\begin{lemma} [S-Procedure~\cite{Convex}]\label{lemma03}
Define a function ${\mathscr{G}}_i\left( {\bf{x}} \right)$, $i \in \{ 1,2\}$, satisfying
\begin{align}
	{\mathscr{G}}_i\left( {\bf{x}} \right)\buildrel \Delta \over = {{\bf{x}}^H}{{\bf{A}}_i}{\bf{x}} + 2{\mathop{{\mathfrak{Re}}}\nolimits} \left\{ {{\bf{b}}_i^H{\bf{x}}} \right\} + {c_i},
	\label{eq27}\end{align}
where ${{\bf{A}}_i} \in \mathbb{H}^{N}$, ${\bf{b}}_i \in \mathbb{C}^{N \times 1}$, and ${c_i} \in \mathbb{R}$. The expression ${\mathscr{G}}_1\left( {\bf{x}} \right) \le 0 \Rightarrow {\mathscr{G}}_2\left( {\bf{x}} \right)\le 0$ holds if and only if there exists $\varsigma>0$ such that
\begin{eqnarray}
	\begin{aligned}[b]
		\varsigma \left[ {\begin{array}{*{20}{c}}
				{{{\bf{A}}_1}}&{{{\bf{b}}_1}}\\
				{{\bf{b}}_1^H}&{{c_1}}
		\end{array}} \right] - \left[ {\begin{array}{*{20}{c}}
				{{{\bf{A}}_2}}&{{{\bf{b}}_2}}\\
				{{\bf{b}}_2^H}&{{c_2}}
		\end{array}} \right]\succeq \boldsymbol{0}.
	\end{aligned}
	\label{eq28}\end{eqnarray}
\end{lemma}
Obviously,~\eqref{eq28} specifies that there exists a vector ${{\bf{\hat x}}}$ satisfying ${\mathscr{G}}_i\left( {\bf{\hat x}} \right)<0$. We rewrite the uncertainty set in~\eqref{eq25} as
\begin{align}
	{\mathscr{G}}_1\left(\Delta {{\bf{h}}_{k}}\right)=\Delta {{\bf{h}}_{k}^H}\Delta {{\bf{h}}_{k}} - \varepsilon _k^2 \le 0, \forall k \in \mathcal{K}.
	\label{eq29}\end{align}
Then, inserting ${{\bf{h}}_{k}} = {{\bf{\hat h}}_{k}} + \Delta {{\bf{h}}_{k}}$ into constraint~\eqref{eq26} and applying Lemma~\ref{lemma03}, we obtain
\begin{align}
	{\mathscr{G}}_2\left(\Delta {{\bf{h}}_{k}^H}\right)&=\Delta {\bf{h}}_{k}^H\left({{\bf{W}}_k} - \gamma \sum\limits_{i \in {\cal K}\backslash \{ k\} } {{{\bf{W}}_i}}  \right)\Delta {{{\bf{h}}}_{k}} \notag\\&
	+ 2{\mathop{{\mathfrak{Re}}}\nolimits} \left\{ {{\bf{\hat h}}_{k}^H\left({{\bf{W}}_k} - \gamma \sum\limits_{i \in{\cal K}\backslash \{ k\} } {{{\bf{W}}_i}}  \right)\Delta {\bf{h}}_{k}} \right\} \notag\\&
	+ {\bf{\hat h}}_{k}^H\left({{\bf{W}}_k} - \gamma \sum\limits_{i \in {\cal K}\backslash \{ k\} } {{{\bf{W}}_i}} \right){{\bf{\hat h}}_{k}} - \sigma^2, \forall k \in \mathcal{K}.
	\label{eq30}\end{align}
To conduct ${\mathscr{G}}_2(\Delta {{\bf{h}}_{k}}) \le 0$, there exist slack variables ${\varsigma _k} \ge 0$, such that
\begin{align}
	\begin{array}{l}
		{{\bf{S}}_r}\left( {{{\bf{W}}_k},{\varsigma _k}} \right)\\
		= \left[ {\begin{array}{*{20}{c}}
				{{\varsigma _k}{{\bf{I}}_N}}&{0}\\
				{0}&{- {\varsigma _k}\varepsilon _k^2 - \gamma\sigma^2}
		\end{array}} \right]\\
		+ {\bf{U}}_{k}^H\left({{\bf{W}}_k} - \gamma \sum\limits_{i \in{\cal K}\backslash \{ k\} } {{{\bf{W}}_i}} \right) {{\bf{U}}_{k}} \succeq \boldsymbol{0}, \forall k \in  \mathcal{K},
	\end{array}
	\label{eq31}\end{align}
where ${{\bf{U}}_{k}} = [ {{{\bf{I}}_N},{{\bf{\hat h}}}_{k}} ]$. It can be easily found that the non-convex constraint has been transformed to a finite number of convex one, which is conducive to obtaining the optimal solutions.

\subsection{High-Dimensional Virtual Array}
\begin{figure}[t]
	\begin{centering}
		\subfigure[]{\begin{centering}
				\includegraphics[width=0.5\columnwidth]{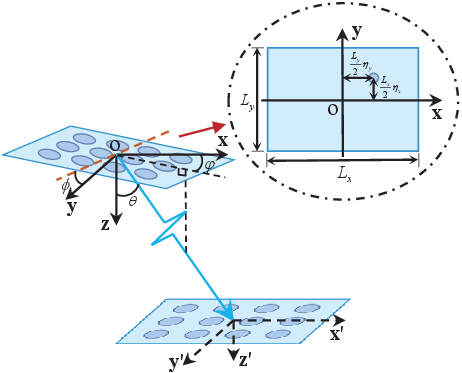}
				\par\end{centering}
		}\subfigure[]{\begin{centering}
				\includegraphics[width=0.42\columnwidth]{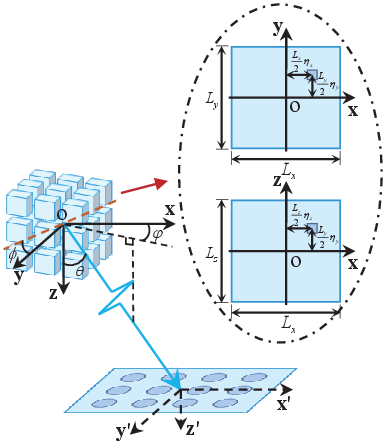}
				\par\end{centering}
		}
		\par\end{centering}
	\caption{Model of high dimensional virtual non-uniform array. (a) 2D planar array. (b) 3D cube array.}
	\label{fig3}\end{figure}
To show the transmission characteristics, we focused on the virtual linear array, previously. Actually, the virtual array performed by UAVs is easily constructing multidimensional arrays, which have a much higher orientation flexibility and a much better power concentration~\cite{Structures}.

Next, we extend the previous scheme to the systems with 2D non-uniform planar array (NUPA) and 3D non-uniform cube array (NUCA) by treating each axes as a virtual NULA. The virtual NUPA (or NUCA) consists of $N$ UAVs, i.e., $N = N_x \times N_y  (\times N_z)$ with aperture sizes $L_x$ and $L_y$ ($L_z$). The normalized spacing vectors along two (three) axes are denoted by ${{\boldsymbol{\eta }}_x}=\{{\eta }_{x,n_x}\}\in \mathbb{R} ^{N_x \times 1}$ and ${{\boldsymbol{\eta }}_y}=\{{\eta }_{y,n_y}\}\in \mathbb{R} ^{N_y \times 1}$ (${{\boldsymbol{\eta }}_z}=\{{\eta }_{z,n_z}\}\in \mathbb{R} ^{N_z \times 1}$), respectively. The BS is equipped with a UPA, which consists of $M$ antennas, i.e., $M = M_x \times M_y$. Similarly, by employing 3D geometrical coordinates, as illustrated in Fig. \ref{fig3}, we model the far-field LoS MIMO channel. The origin is set as the center of virtual NUPA (or NUCA). The $x$-axis and $y$-axis are parallel to columns and rows of the received UPA on the ground, respectively. The $z$-axis points to ground. The rotation offset rotates around $z$-axis. Then, the specific coordinates of each transmit virtual NUPA (or NUCA)/received element are determined as
\begin{align}
\left\{ {\begin{array}{*{20}{l}}
 		{{x_{U,n_x,n_y}} = \frac{{{L_x}{\eta _{x,n_x}}\cos \phi  - {L_y}{\eta _{y,n_y}}\sin \phi }}{2}},\\
 		{{y_{U,n_x,n_y}} = \frac{{{L_x}{\eta _{x,n_x}}\sin \phi  + {L_y}{\eta _{y,n_y}}\cos \phi }}{2}},\\
 		{z_{U,n_z} = 0,(z_{U,n_z}=\frac{{{L}{\eta _{n_z}}}}{2})},
 \end{array}} \right.
	\label{eq32}\end{align}
and 
\begin{align} 
\left\{ {\begin{array}{*{20}{l}}
		{{x_{B,{m_x}}} = \frac{{2{m_x} - 1 - {M_x}}}{2}d_x + R\sin \theta \cos \varphi ,}\\
		{{y_{B,{m_y}}} = \frac{{2{m_y} - 1 - {M_y}}}{2}d_y + R\sin \theta \sin \varphi ,}\\
		{{z_{B,{m_z}}} = R\cos \theta .}
\end{array}} \right.
	\label{eq33}\end{align}

Following similar steps as virtual NULA, we construct an asymptotic channel matrix by calculating the specific wave propagation range ${\tau_{m,n}}$ between the $n$th transmit UAV and the $m$th received element, and then obtain the optimal UAV topology followed the 2D/3D Fekete distributions. Next, a similar minimizing total transmit power problem is formulated to design the trajectory and precoding.

\section{Simulation Results}
\begin{figure}[tb]
	\begin{center}
		\includegraphics[width=1\columnwidth]{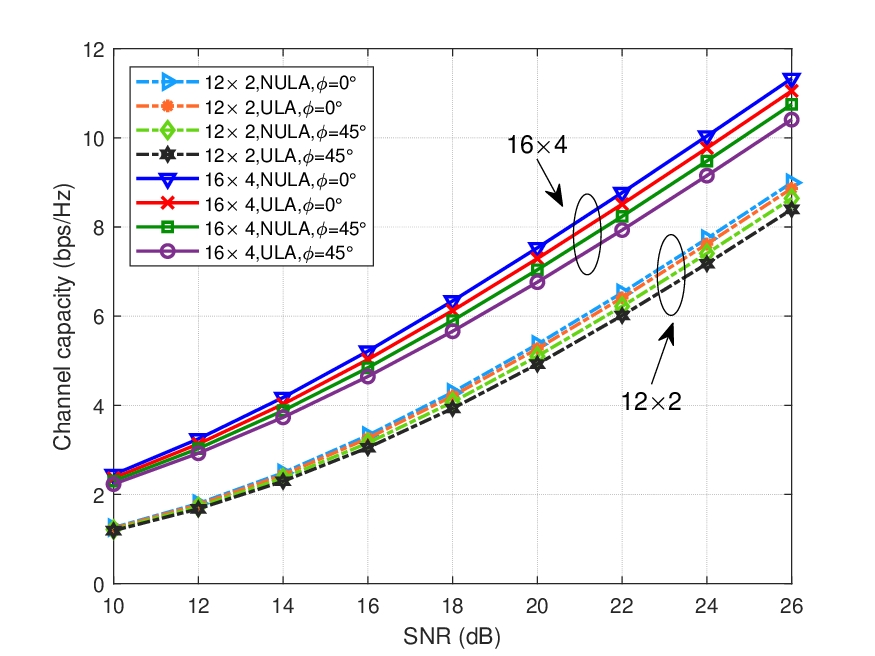}
	\end{center}
		\caption{Channel capacity versus prescribed minimum SNR, for different array scales and rotation offsets.}
	\label{fig4}
\end{figure}
\begin{figure}[tb]
	\begin{center}
		\includegraphics[width=1\columnwidth]{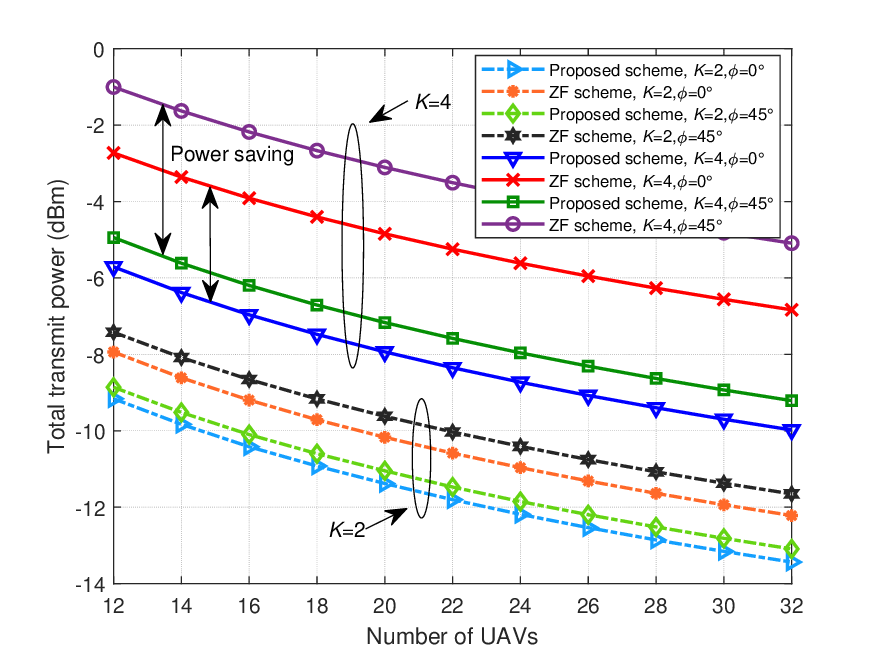}
	\end{center}
	\caption{Total transmit power versus number of UAVs, for different numbers of data streams and rotation offsets.}
	\label{fig5}
\end{figure} 
\begin{figure}[t]
	\begin{centering}
		\subfigure[]{\begin{centering}
				\includegraphics[width=1\columnwidth]{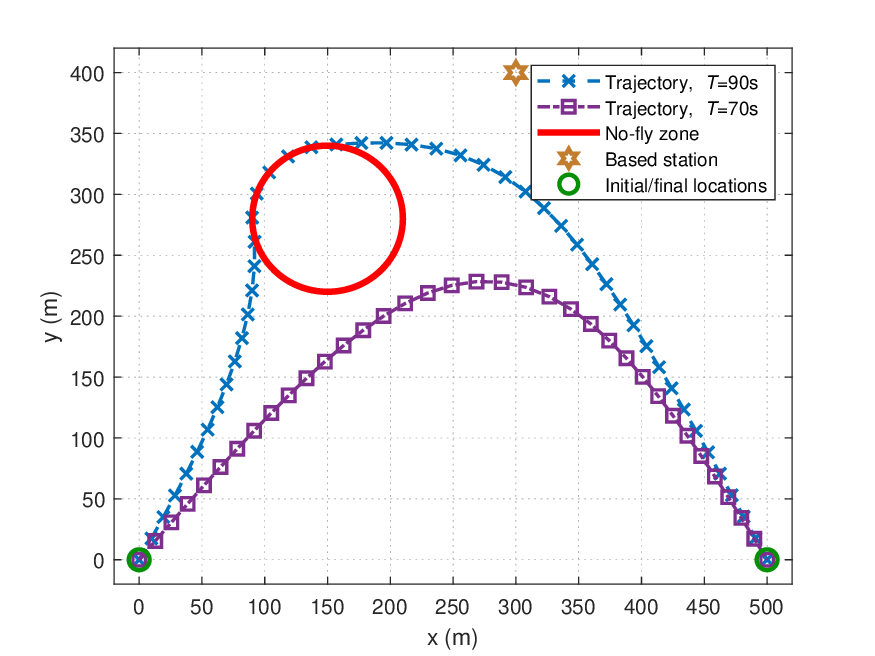}
				\par\end{centering}
		}\\
		\subfigure[]{\begin{centering}
				\includegraphics[width=1\columnwidth]{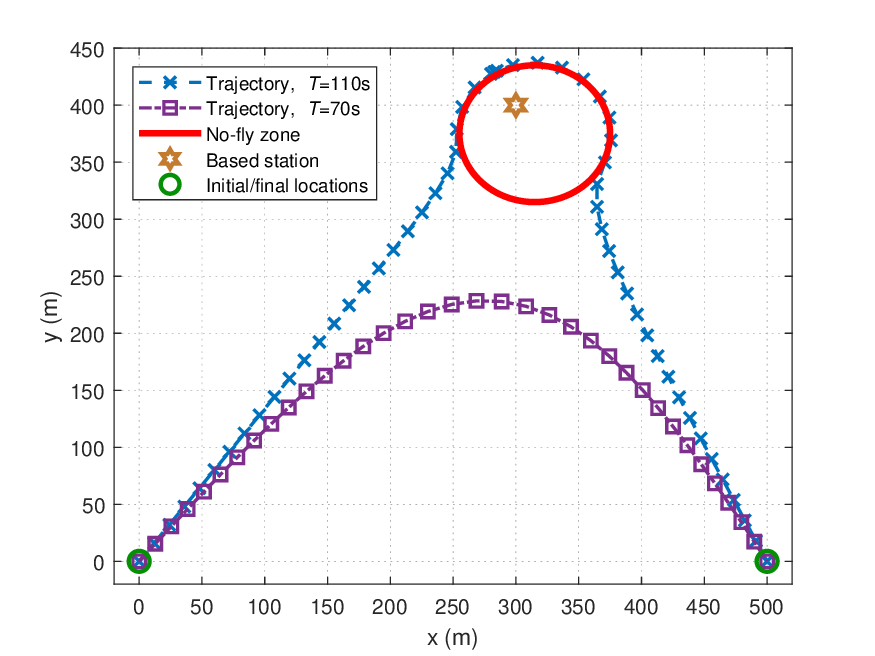}
				\par\end{centering}
		}
		\par\end{centering}
	\caption{Optimal UAV trajectories with different time duration. (a) BS is outside no-fly zone. (b) BS is inside no-fly zone.}
	\label{fig6}\end{figure}
	
In this section, we investigate the performance of the proposed UAV-enabled communication scheme via numerical results. Specifically, the carrier frequency is $f_c=1 {\rm ~GHz}$. The initial and final locations in 2D area are fixed as ${\bf d}_I=(0{\rm m},0{\rm m})$ and ${\bf d}_F=(500{\rm ~m},0{\rm ~m})$ with constant altitude of $100{\rm ~m}$. The center coordinate of the BS is $(300{\rm ~m}, 400{\rm ~m})$. The maximum transmit power of each UAV is set to $P_{\rm max} = 10 {\rm ~dBm}$. All UAV’s maximum flight speed is $V_{\rm max} = 10 {\rm ~m/s}$. The prescribed minimum SNR is set as $\gamma=14 {\rm ~dB}$. The transmit virtual NULA aperture size is set as $L=10 {\rm ~m}$. For simplicity, we assume $10\lg (\sigma^2)=-110$ \text{dBm} \cite{Robust}. The number of Eves is assumed as $Q=3$. The maximum leakage SNR tolerance of passive Eves is set to $\xi=0 {\rm~dB}$. The probabilistic parameter is chosen as $\kappa = 0.99$. To show the behavior of our proposed UAV topology and precoding design method, we consider two baseline schemes for comparison, namely, the 'ULA' and 'zero forcing (ZF) scheme'~\cite{ZF}.

Figure~\ref{fig4} reveals that the system channel capacities of our proposed NULA and ULA schemes, when the communication range is set as $R=300 {\rm~m}$. We observe the slope difference between the curves with ULA and the proposed NULA schemes under different array scales, especially in high SNR regime. This means that a higher channel capacity gain can be obtained by our proposed NULA scheme. Besides, deploying more antennas improves the capability in array gains, and thus better performance in terms of channel capacity can be achieved. According to~\eqref{eq12}, we know that the eigenvalues of the channel gain matrix are actually negatively associated with the rotation offsets. Consistent with the results in Fig.~\ref{fig4}, larger rotation offsets yield channel capacity dropping to a certain extent. This is attributed to the fact that the realized channel capacity is corresponding to the projected equivalent array aperture.
\begin{figure}[tb]
	\begin{center}
		\includegraphics[width=1\columnwidth]{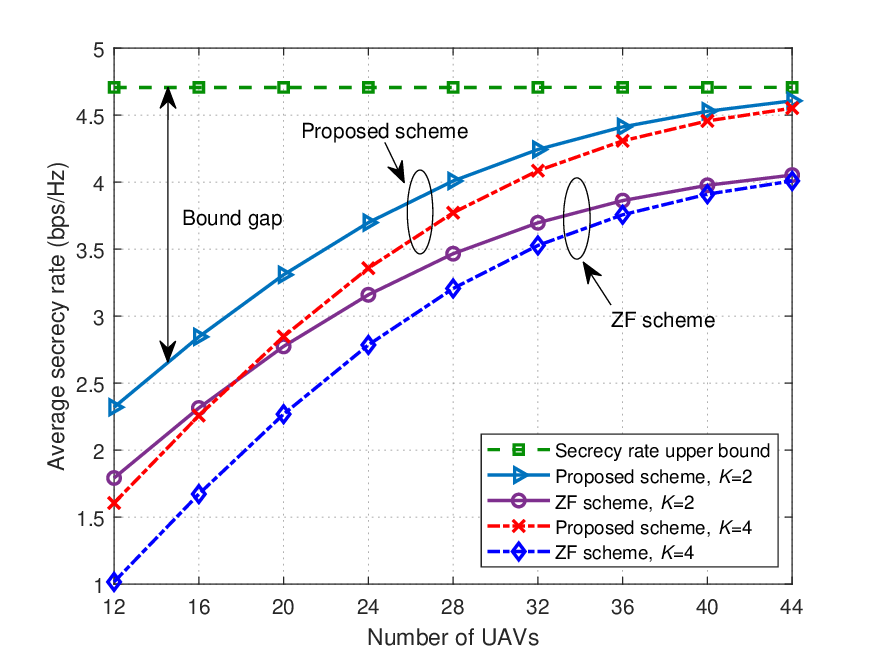}
	\end{center}
	\caption{Average secrecy rate versus number of UAVs, for different numbers of data streams.}
	\label{fig7}
\end{figure}

In Fig.~\ref{fig5}, we illustrate how the number of UAVs influences the total transmit power consumption. It can be observed that the total transmit power decreases with the increase of the number of UAVs. The reason is that the virtual array deployed with more UAVs is capable of more efficient design. A general observation is that lower transmit power consumption can be achieved for our proposed joint design method than conventional ZF method under the same conditions, which demonstrates more power saving in improving system energy efficiency. The gap between the two schemes is more pronounced for a large $K$. Interestingly, the introduction of rotation offset requires the consumption of extra power. Obviously, there is a trade off between secrecy performance and rotation offsets. Larger rotation offset leads to greater dynamic changes, yet more transmit power consumption. Therefore, we should balance the secure performance improvement and the power consumption in practice.
\begin{figure}[tb]
	\begin{center}
		\includegraphics[width=1\columnwidth]{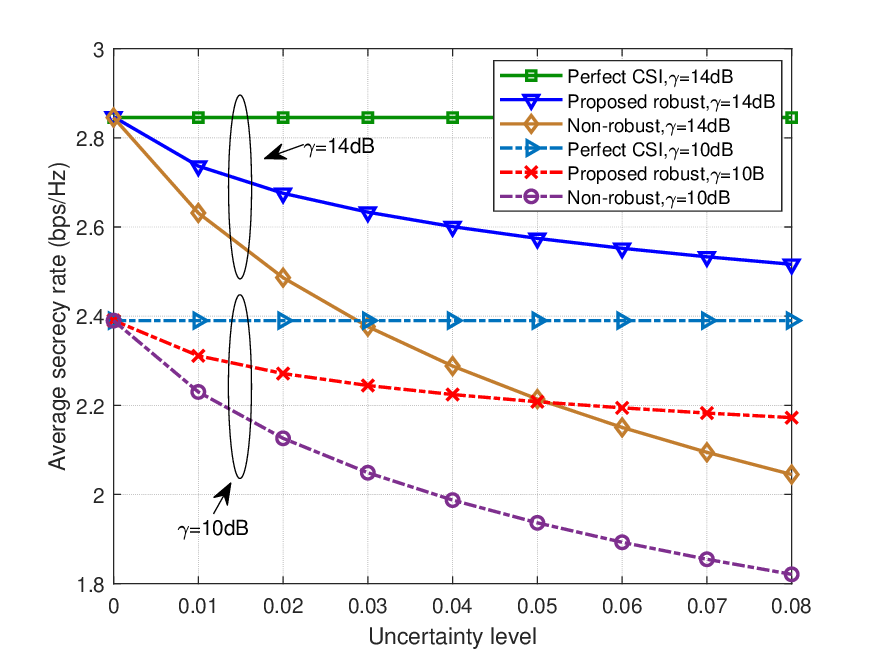}
	\end{center}
	\caption{Average secrecy rate versus uncertainty level, for different receive SNRs.}
	\label{fig8}
\end{figure}

\begin{figure}[t]
	\begin{centering}
		\subfigure[]{\begin{centering}
				\includegraphics[width=1\columnwidth]{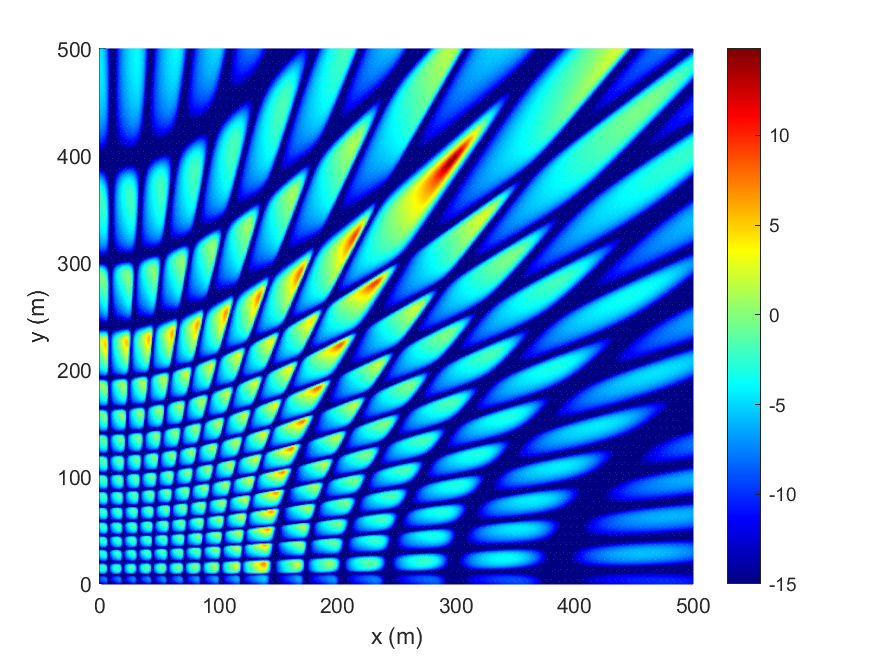}
				\par\end{centering}
		}\\
		\subfigure[]{\begin{centering}
				\includegraphics[width=1\columnwidth]{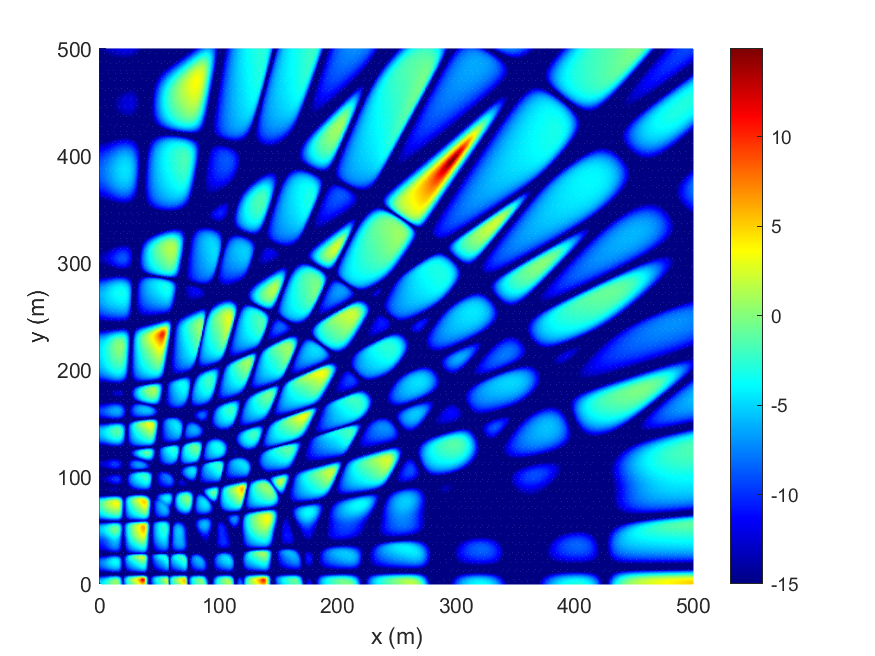}
				\par\end{centering}
		}
		\par\end{centering}
	\caption{SNR radiation of virtual array versus coordinate on the ground. (a) 2D planar array. (b) 3D cube array.}
	\label{fig9}\end{figure}
	
To show the behavior of no-fly zone in practice, we consider two scenarios that the center coordinate of the BS is set outside no-fly zone, i.e., ${\boldsymbol \chi }=(150 {\rm~m}, 280 {\rm ~m})$, and inside no-fly zone, i.e., ${\boldsymbol \chi }=(315 {\rm ~m}, 375 {\rm ~m})$, with radius $\Upsilon=60 {\rm ~m}$. Fig.~\ref{fig6} illustrates the UAV trajectory with three different total flight duration, $T =70{\rm ~s}$, $T =90{\rm ~s}$, and $T =110{\rm ~s}$, respectively. It is observed that when the total flight duration is relatively small (e.g. $T = 70{\rm ~s}$), UAVs approach the BS and hover over it, and then they return to the destination. In addition, when the flight duration is sufficiently long  (e.g. $T = 90{\rm ~s}$ and $T =110{\rm ~s}$), we can see that the UAVs modify their behavior to fully benefit from the additional time affords. In particular, when the flight direction between the UAVs and the BS is blocked by the no-fly zone, the optimal UAV trajectory would bypass the no-fly zone and follow the tangential line of the no-fly zone to provide the ability to have a shorter flight path. Meanwhile, the UAVs hover over for prolonged periods of time in locations near the BS, which conforms to the target of minimum power consumption policy.

Generally, the average secrecy rate is viewed as a main metric on the secrecy performance of the transmission systems~\cite{passiveEve}. Then, the average secrecy rate for different numbers of UAVs is illustrated. The Eve is picked up over $10^2$ random samples. For a fair comparison, both schemes are set as the same total transmit power. According to the definition of secrecy rate, the secrecy rate upper bound is also introduced, i.e., under no Eve existence. As can be seen from Fig.~\ref{fig7}, increasing the number of UAVs improves the average secrecy rate. Again, our proposed scheme is superior to the conventional ZF scheme. This result is in light of the fact that our careful design of the UAV topology, trajectory, and precoding help improve the channel capacity and reduce information leakage.

Fig.~\ref{fig8} depicts the average secrecy rate versus uncertainty levels. For convenience, we unify the radius  as $\varepsilon=\varepsilon_m, \forall m$ and define the normalized uncertainty levels as $ \varepsilon^2 / \rho^2 (R),\forall k $. As seen, the non-robust method cannot obtain the optimal secure performance  due to the CSI uncertainty, and the average secrecy rate decreases as the uncertainty level increases. As a contrast, the performance of our proposed robust approach degrades to a certain extent as the uncertainty level increases, but can achieve a higher average secrecy rate than non-robust method. 
	
Finally, we investigate the spatial distribution of high-dimensional virtual arrays on the performance in Fig.~\ref{fig9}, i.e., 2D NUPA and 3D NUCA. Concurrently, to achieve fair comparison, the number of transmit arrays are both chosen to $N=64$. As expected, two types of arrays show similar radiation patterns, and the received SNR at the BS is ${\gamma}=14{\rm~dB}$. That is, our designs meet to the QoS requirements. At the same time, 3D cube array provides effective SNR suppresses along the sidelobes due to a more flexible antenna array structure. 
	
\section{Conclusion}
In this paper, we studied the multi-UAV cooperative PHY security problem, under which many conventional array approaches fail to deploy in UAV practical scenarios due to size and power limitations of a single UAV. To tackle this issue, we used a virtual NULA formed by a cluster of UAVs to enhance transmission performance. Specifically, we aimed to maximize the system channel capacity by jointly optimizing the topology, precoding, and trajectory. To solve this challenging problem, some insights of the solutions were given. Afterwards, we divided the origin problem into two independent problems. The channel gain was improved by designing the UAV topology, and the freedom of space was enhanced via randomizing rotation offsets. On the other hand, through the design of precoding and trajectory, the total transmission power was minimized, while meeting the QoS requirements, the leakage SNR tolerance, the single UAV transmit power constraint, and some trajectory constraints. Moreover, we provided extensions of the proposed scheme to the cases of imperfect CSI and high-dimensional virtual arrays. Finally, the superiority of proposed multi-UAV cooperation-based PHY security scheme was validated via simulation results. The proposed scheme can be widely applied foreground to the near future. The virtual array formed by UAVs to collaborative beamforming is an emerging technology. By leveraging the collective power, this work opens a way to flexible communication with significant enhancement of capabilities, coverage, and signal strength. Some potential application scenarios include emergency communications, coverage of remote areas, and military applications. 

Due to the space limitations, there are various interesting challenges unaddressed in this paper, which are briefly discussed in the following to motivate future work. The strict synchronization strategy among the UAVs and the performance degradation caused by environmental impacts are both worthwhile works.

\appendices
\section{Proof of Theorem 1}
Let us perform Taylor expansion on \eqref{eq10}, i.e.,
\begin{eqnarray}
	\begin{aligned}[b]
	{{\tilde h}_{m,n}}& = {e^{j\nu \cos \phi {\eta _n}(2m - 1 - M)/M}}\\&
	 = \sum\limits_{\alpha = 0}^\infty  {\frac{[j\nu \cos \phi {\eta _n}(2m - 1 - M)/M]^\alpha }{\alpha!}}.
\end{aligned}
\label{eq34}\end{eqnarray}
Utilizing \eqref{eq34}, ${\bf{\tilde H}}$ can be decomposed as
\begin{align}
{\bf{\tilde H}} = {{\bf{V}}_B}{\bf{A}}{\bf{V}}_U^T,
	\label{eq35}\end{align}
where ${\bf{A}}={\rm diag}(a_1,a_2,...)$ is a $\infty$-by-$\infty$ diagonal matrix with ${a_\alpha} = {(j\nu \cos \phi )^{\alpha - 1}}/(\alpha - 1)!, \alpha \in [1,2,...]$. ${{\bf{V}}_U} \in {{\mathbb R}^{N \times \infty }}$ and ${{\bf{V}}_B} \in {{\mathbb R}^{M \times \infty}}$ indicate the Vandermonde matrices related to the transmit-receive topology, which are given as
\begin{align}
{{\bf{V}}_U} = {\left( {\begin{array}{*{20}{c}}
			1&{{\eta _1}}&{\eta _1^2}& \cdots \\
			1&{{\eta _2}}&{\eta _2^2}& \cdots \\
			\vdots & \vdots & \vdots & \ddots \\
			1&{{\eta _N}}&{\eta _N^2}& \ldots 
	\end{array}} \right)_{N \times \infty }},
	\label{eq36}\end{align}
and 
\begin{align}
{{\bf{V}}_B} = {\left( {\begin{array}{*{20}{c}}
			1&{\frac{{1 - M}}{M}}&{{{(\frac{{1 - M}}{M})}^2}}& \cdots \\
			1&{\frac{{3 - M}}{M}}&{{{(\frac{{3 - M}}{M})}^2}}& \cdots \\
			\vdots & \vdots & \vdots & \ddots \\
			1&{\frac{{M - 1}}{M}}&{{{(\frac{{M - 1}}{M})}^2}}& \ldots 
	\end{array}} \right)_{M \times \infty }}.
	\label{eq37}\end{align}
We decompose the matrices ${{\bf{V}}_B}$, ${{\bf{A}}}$, ${{\bf{V}}_U}$ into sub-matrices as ${{\bf{V}}_B}=({{\bf{V}}_{B,1}},{{\bf{V}}_{B,2}},...)$, ${\bf{A}}={\rm diag}({\bf{A}}_{1},{\bf{A}}_{2},...)$, and ${{\bf{V}}_U}=({{\bf{V}}_{U,1}},{{\bf{V}}_{U,2}},...)$, where ${{\bf{V}}_{B,i}}$,${{\bf{V}}_{U,i}}$, and ${{\bf{A}}_{i}}$ are the $i$th $M$-by-$M$ sub-matrix of ${{\bf{V}}_B}$, $N$-by-$M$ sub-matrix of ${{\bf{V}}_U}$, and $M$-by-$M$ diagonal sub-matrix of ${{\bf{A}}}$, respectively. Then, \eqref{eq35} can be equivalently expressed as
\begin{align}
{\bf{\tilde H}} = \sum\limits_{i = 1}^\infty  {{{\bf{\tilde H}}}_i},
	\label{eq38}\end{align}
where ${{\bf{\tilde H}}_i} = {{\bf{V}}_{B,i}}{{\bf{A}}_i}{\bf{V}}_{U,i}^T$. Next, we focus on the first sub-matrix ${{\bf{\tilde H}}_1}$ for approximation. Let ${\tilde \lambda _m }$ indicate the $m$th eigenvalue of ${{\bf{\tilde \Pi }}_1} \buildrel \Delta \over = {{\bf{\tilde H}}_1}{\bf{\tilde H}}_1^H$, and we get
\begin{align}
	&\mathop {\lim }\limits_{\nu  \to 0} \frac{{\ln \left( {\prod\limits_{m \in {\cal M}} {\tilde \lambda }_m}  \right)}}{{\ln \left( {\nu \cos \phi } \right)}}\nonumber\\&
	\!	=\! \mathop {\lim }\limits_{\nu  \to 0} \frac{{\ln \left( {\det ({{{\bf{\tilde \Pi }}}_1})} \right)}}{{\ln ({\nu \cos \phi })}}\nonumber\\&
	\!	=\! \mathop {\lim }\limits_{\nu  \to 0} \frac{{\ln \left( {\det ( {{{\bf{V}}_{B,1}}{{\bf{A}}_1}{\bf{V}}_{U,1}^T{{\bf{V}}_{U,1}}{\bf{A}}_1^H{\bf{V}}_{B,1}^T})} \right)}}{{\ln \left( {\nu \cos \phi } \right)}}\nonumber\\&
	\!	=\! \mathop {\lim }\limits_{\nu  \to 0} \frac{{\ln \left( {\det ( {{\bf{V}}_{B,1}}{\bf{V}}_{B,1}^T)\det({{\bf{A}}_1})\det({\bf{V}}_{U,1}^T{{\bf{V}}_{U,1}})\det({\bf{A}}_1^H)} \right)}}{{\ln \left( {\nu \cos \phi } \right)}}\nonumber\\&
	\!	=\! \mathop {\lim }\limits_{\nu  \to 0} \frac{{\ln \left( {\det ( {{\bf{V}}_{B,1}^T{{\bf{V}}_{B,1}}}   {{\bf{V}}_{U,1}^T{{\bf{V}}_{U,1}}}) \frac{{{{\left( {\nu \cos \phi } \right)}^{M(M\! -\! 1)}}}}{{{{\left[{\prod\limits_{m \in {\cal M}} {\left( {m\! -\! 1} \right)!} }\right]}^2}}}} \right)}}{{\ln \left( {\nu \cos \phi } \right)}}\nonumber\\&
	\!	=\! M(M - 1).
\label{eq39}\end{align}

Next, the Vandermonde matrix ${{\bf{V}}_{B,1}}$ is operated with QR decomposition, i.e.,
\begin{align}
{{\bf{V}}_{B,1}} = {{\bf{Q}}_{B,1}}{{\bf{R}}_{B,1}},
	\label{eq40}\end{align}
where ${{\bf{Q}}_{B,1}}\in {\mathbb{R}^{M \times M}}$ denotes the unitary matrix and ${{\bf{R}}_{B,1}}\in {\mathbb{R}^{M \times M}}$ denotes the upper triangular matrix. Let's construct a matrix as
\begin{align}
	{\bf G} \buildrel \Delta \over  = {\bf{Q}}_{B,1}^T{\tilde {\bf \Pi}_1} {{\bf{Q}}_{B,1}}
	= {{\bf{R}}_{B,1}}{{\bf{A}}_1}{\bf{V}}_{U,1}^T{{\bf{V}}_{U,1}}{\bf{A}}_1^T{{\bf R}}_{B,1}^T.
	\label{eq41}\end{align}
It is easy to see that ${\bf G}$ has the same eigenvalue as ${\tilde {\bf \Pi}_1}$. We now consider the $m$th diagonal entry of ${\bf G}$, denoted by ${{g}_m}$, satisfying~\cite{Uniform}
\begin{align}
\mathop {\lim }\limits_{\nu  \to 0} \frac{{\ln {{g}_m}}}{{\ln \left( {\nu \cos \phi } \right)}} = 2(m - 1), \forall m \in {\cal M}.
	\label{eq42}\end{align}
When $\nu \to 0$, we have ${\ln (v\cos \phi )}\to - \infty $. The bound the asymptotic slope is given by
\begin{align}
	\mathop {\lim }\limits_{\nu  \to 0} \frac{{\ln \left( {{\tilde \lambda }_m} \right)}}{{\ln \left( {\nu \cos \phi } \right)}} \! \ge \!\mathop {\lim }\limits_{\nu  \to 0} \frac{{\ln \left( {\sum\limits_{i \in  {\cal M}_m} {\tilde \lambda }_i} \right)}}{{\ln \left( {\nu \cos \phi } \right)}}\!\overset{{(a)}}\ge \! \mathop {\lim }\limits_{\nu  \to 0} \frac{{\ln \left( {\sum\limits_{i \in  {\cal M}_m} {{g_i}} } \right)}}{{\ln \left( {\nu \cos \phi } \right)}},
\label{eq43}\end{align}
where ${{\cal M}_m}\buildrel \Delta \over=[m,m+1,...,M]$. The inequality transformation $(a)$ can be got due to the Hermitian matrix majorization relations~\cite[Co. 4.3.34.]{Matrix}. Then, we have
\begin{align}
\mathop {\lim }\limits_{\nu  \to 0} \frac{{\ln \left( {\prod\limits_{m \in {\cal M}} {\tilde \lambda }_m} \right)}}{\ln \left( {\nu \cos \phi } \right)}\! = \!\mathop {\lim }\limits_{\nu  \to 0} \sum\limits_{m \in {\cal M}} {\frac{\ln \left( {{\tilde \lambda }_m} \right)}{\ln \left( {\nu \cos \phi } \right)}} \! \ge\! M(M \!- \!1).
\label{eq44}\end{align}
Combining~\eqref{eq39},~\eqref{eq43}, and~\eqref{eq44}, we have
\begin{align}
	\mathop {\lim }\limits_{\nu  \to 0} \frac{{\ln \left( {{\tilde \lambda }_m} \right)}}{\ln \left( {\nu \cos \phi } \right)} = 2(m - 1),  \forall m \in {\cal M}.
		\label{eq45}\end{align}

From~\eqref{eq38}, we can regard ${\bf{\tilde H}}$ as a perturbation form of ${{{\bf{\tilde H}}}_1}$. Utilizing the Weyl’s Perturbation Theorem~\cite{Perturbation,Weyl}, we get the relationship between ${{\tilde \lambda }_m}$ and ${{\lambda }_m}$ as
\begin{align}
	| {\sqrt {\lambda _m}  - \sqrt {\tilde \lambda }_m} | \le \| {\sum\limits_{i = 2}^\infty  {{\bf{\tilde H}}}_i} \|_2.
		\label{eq46}\end{align}
The right-side of~\eqref{eq46} can be modified to
\begin{eqnarray}
	\begin{aligned}[b]
	&\mathop {\lim }\limits_{\nu  \to 0} \frac{\| {\sum\limits_{i = 2}^\infty  {{\bf{\tilde H}}}_i} \|_2}{{( \nu \cos \phi )}^{M - 1}}\\&
	\le \mathop {\lim }\limits_{\nu  \to 0} \frac{\sum\limits_{i = 2}^\infty  {\| {\bf{V}}_{B,i} \|_2}  \cdot \| {{\bf{A}}_i} \|_2 \cdot \| {{\bf{V}}_{U,i}^T} \|_2}{{( \nu \cos \phi )}^{M - 1}}\\&
	\overset{{(b)}}= \mathop {\lim }\limits_{\nu  \to 0} \frac{\sum\limits_{i = 2}^\infty  {\| {\bf{V}}_{B,i} \|_2}  \cdot \frac{{{( \nu \cos \phi )} ^{M(i - 1)}}}{[ {M(i - 1)}]!} \cdot \| {{\bf{V}}_{U,i}^T} \|_2}{{( \nu \cos \phi )}^{M - 1}}= 0.
\end{aligned}
\label{eq47}\end{eqnarray}
The equality $(b)$ is obtained since spectral norm of a matrix is the largest singular value~\cite[Th. 5.6.2.]{Matrix}. Based on~\eqref{eq46} and~\eqref{eq47}, we have
\begin{eqnarray}
	\begin{aligned}[b]
	&\mathop {\lim }\limits_{\nu \to 0} | {\sqrt {{\lambda _m}/{{(\nu \cos \phi)}^{2(m - 1)}}}  - \sqrt {{{\tilde \lambda }_m}/{{(\nu \cos \phi)}^{2(m - 1)}}} } |\\&
	\le \mathop {\lim }\limits_{\nu  \to 0} \frac{{\| {\sum\limits_{i = 2}^\infty  {{\bf{\tilde H}}}_i} \|_2}}{{\left( {\nu \cos \phi } \right)}^{M - 1}}= 0,
\end{aligned}
\label{eq48}\end{eqnarray}
which implies that ${\lambda _m}$ and ${{\tilde \lambda }_m}$ are the high-order infinitesimal term with ${(\nu \cos \phi)}^{2(m - 1)}$. As a result, we can draw the conclusion as
\begin{align}
	\mathop {\lim }\limits_{\nu  \to 0} \frac{{\ln ( \lambda _m )}}{\ln ( \nu \cos \phi )} = \mathop {\lim }\limits_{\nu  \to 0} \frac{{\ln ( {{{\tilde \lambda }_m}} )}}{\ln (\nu \cos \phi )} = 2(m - 1), \forall m \in {\cal M}.
\label{eq49}\end{align}

Denote by $\{{\boldsymbol{\zeta }}_m\} $ the eigenvector of  ${\bf{\Pi }}$ corresponding to eigenvalue $\{{\lambda _m}\}$, which  span a basis of vector space. Thus, each column of ${{\bf{Q}}_{B,1}}$ is a linear combination of all the eigenvectors, i.e.,
\begin{align}
{{\bf{q}}_{B,m}} = \sum\limits_{i \in {\cal M}} {{\delta _{m,i}}{{\boldsymbol{\zeta }}_i}},\forall m \in {\cal M},
	\label{eq50}\end{align}
where ${{\bf{q}}_{B,m}}$ is the $m$th column of ${{\bf{Q}}_{B,1}}$, and $\{\delta _{m,i}\}$ are real numbers and satisfy  $\sum\nolimits_{i \in {\cal M}} {\delta _{m,i}^2}  = 1,\forall m$.

Assume ${{\bf{q}}_{B,M}} \ne {{\boldsymbol{\zeta }}_M}$, as $\nu  \to 0$. Then we have ${\delta _{M,M}} =0$ and $\exists {i^{\star}}({i^*} < M)$ so as to ${\delta _{M,{i^*}}}\ne0$, yielding
\begin{eqnarray}
	\begin{aligned}[b]
	\mathop {\lim }\limits_{\nu  \to 0} {\lambda_M}& = \mathop {\lim }\limits_{\nu  \to 0} {\bf{q}}_{B,M}^T{\bf{\Pi }}{{\bf{q}}_{B,M}}\\&
	= \mathop {\lim }\limits_{\nu  \to 0} {\left( {\sum\limits_{i \in {\cal M}} {{\delta _{M,i}}{{\boldsymbol{\zeta }}_i}} } \right)^T}{\bf{\Pi }}\left( {\sum\limits_{i \in {\cal M}} {{\delta _{M,i}}{{\boldsymbol{\zeta }}_i}} } \right)\\&
	= \mathop {\lim }\limits_{\nu  \to 0} \sum\limits_{i \in {\cal M}} {\delta _{M,i}^2} {\boldsymbol{\zeta }}_i^T{\bf{\Pi }}{{\boldsymbol{\zeta }}_i}\\&
	= \mathop {\lim }\limits_{\nu  \to 0} \sum\limits_{i \in {\cal M}} {\delta _{M,i}^2} {\lambda _i} \ge \delta _{M,{i^*}}^2{\lambda _{{i^*}}}.
\end{aligned}
\label{eq51}\end{eqnarray}
This leads to
\begin{align}
	\mathop {\lim }\limits_{\nu  \to 0} \frac{\ln ({\lambda_M})}{\ln (v\cos \phi )} \!\le \!\mathop {\lim }\limits_{\nu  \to 0} \frac{\ln (\delta _{M,{i^*}}^2{\lambda _{i^*}})}{\ln (v\cos \phi )}\! =\! 2({i^*} \!-\! 1) < 2(M \!-\! 1).
\label{eq52}\end{align}
Obviously,~\eqref{eq52} does not match the conclusion~\eqref{eq49}. Hence the assumption fails, and we have ${{\bf{q}}_{B,M}} = {{\boldsymbol{\zeta }}_M}$.

Considering the facts ${{\boldsymbol{q}}_{B,M}} = {{\boldsymbol{\zeta }}_M}$ and ${{\boldsymbol{\zeta }}_M} \bot {{\boldsymbol{\zeta }}_{M - 1}}$, we have ${\delta _{M-1,M}} = 0$. Similarly, assuming ${{\bf{q}}_{B,M-1}} \ne {{\boldsymbol{\zeta }}_{M-1}}$ for the case of $i=M-1$,  we have
\begin{align}
	\mathop {\lim }\limits_{\nu  \to 0} \frac{\ln ({\lambda_{M-1}})}{\ln (v\cos \phi )} < 2(M - 2).
	\label{eq53}\end{align}
This inequality is still contrary to the conclusion~\eqref{eq49}. By recursive argument, we can draw the conclusion as follows
\begin{align}
{{\bf{q}}_{B,m}} = {{\boldsymbol{\zeta }}_{m}}, \forall m \in {\cal M}.
	\label{eq54}\end{align}
	
It is worth noting that the left singular vectors of ${\bf{\tilde H}}$ converge to the $\{{\bf{q}}_{B,m}\}$ as ${\nu  \to 0}$. In a similar way, the right singular vectors of ${\bf{\tilde H}}$ converge to the $\{{\bf{q}}_{U,m}\}$ with ${{\bf{q}}_{U,m}}$ being the $m$th column of ${{\bf{Q}}_{U,1}}$. Then, we have
\begin{eqnarray}
	\begin{aligned}[b]
	\mathop {\lim }\limits_{\nu  \to 0} \frac{{\lambda _m}}{{(\nu \cos \phi )}^{2(m - 1)}}
	= \mathop {\lim }\limits_{\nu  \to 0} \frac{{{| {{\bf{q}}_{B,m}^T{\bf{\tilde H}}{{\bf{q}}_{U,m}}} |}^2}}{{(\nu \cos \phi )}^{2(m - 1)}}.
\end{aligned}
\label{eq55}\end{eqnarray}
We perform QR decomposition of ${{\bf{V}}_B}$ and ${{\bf{V}}_U}$ as
\begin{align}
\left\{{\begin{array}{*{20}{c}}
		{{{\bf{V}}_B} = {{\bf{Q}}_B}{{\bf{R}}_B}},\\
		{{{\bf{V}}_U} = {{\bf{Q}}_U}{{\bf{R}}_U}},
\end{array}}\right.
\label{eq56}\end{align}
where ${\bf{Q}}_B \in {\mathbb{R}^{M \times M}}$ and ${\bf{Q}}_U \in {\mathbb{R}^{N \times N}}$ denote the unitary matrix, respectively, ${\bf{R}}_B \in {\mathbb{R}^{M \times \infty}}$ and ${\bf{R}}_U \in {\mathbb{R}^{N \times \infty }}$ denote the upper triangular matrices, respectively.

Denote by $r_{U^(m,n)}$ and $r_{B,(m,n)}$ the $m$th row and the $n$th column element in ${{\bf{R}}_U}$ and ${{\bf{R}}_B}$, respectively. Inserting~\eqref{eq35} and~\eqref{eq56} into~\eqref{eq55}, we obtain
\begin{eqnarray}
	\begin{aligned}[b]
	&\mathop {\lim }\limits_{\nu  \to 0} \frac{{\lambda _m}}{{(\nu \cos \phi )}^{2(m - 1)}}\\&
	= \mathop {\lim }\limits_{\nu  \to 0} \frac{{{| {{\bf{q}}_{B,m}^T{{\bf{V}}_B}{\bf{AV}}_U^T{{\bf{q}}_{U,m}}} |}^2}}{{(\nu \cos \phi )}^{2(m - 1)}}.\\&
	= \mathop {\lim }\limits_{\nu  \to 0} {\left| {\frac{{{\bf{q}}_{R,m}^T{{\bf{Q}}_B}{{\bf{R}}_B}{\bf{A}}{\bf{R}}_U^T{\bf{Q}}_U^T{{\bf{q}}_{U,m}}}}{{(\nu \cos \phi )}^{m - 1}}} \right|^2}\\&
	= \mathop {\lim }\limits_{\nu  \to 0} {\left| {\frac{{\sum\limits_{n = m}^\infty  {{r_{B,(m,n)}}\frac{{(j\nu \cos \phi )}^{n - 1}}{{(m - 1)!}}{r_{U,(m,n)}}}}}{{(\nu \cos \phi )}^{m - 1}}} \right|^2}\\&
	= {\left[{\frac{{r_{B,m}}{r_{U,m}}}{(m - 1)!}} \right]^2},
\end{aligned}
\label{eq57}\end{eqnarray}
where ${r_{B,m}} = {r_{B,(m,m)}}$, and ${r_{U,m}} = {r_{U,(m,m)}}$. The proof of Theorem~\ref{theorem01} is completed.

\section{Proof of Theorem 2}
According to the operation of QR decomposition, ${r_{U,1}}$ is equal to an all-one vector of length $N$ when $k=1$. Thus~\eqref{eq15} holds. Next, we consider the case of $k>1$. Taking the first $k$ columns of the Vandermonde matrix~\eqref{eq36}, denoted by ${{\bf{V}}_{U,(k)}}$, we have
\begin{eqnarray}
	\begin{aligned}[b]
	\det ({\bf{V}}_{U,(k)}^T{{\bf{V}}_{U,(k)}})& = \det ({\bf R}_{U,(k)}^T{\bf Q}_U^T{{\bf Q}_U}{{\bf R}_{U,(k)}})\\&
	= \det ({\bf R}_{U,(k)}^T{{\bf R}_{U,(k)}}) \\&
	\overset{{(c)}} = \prod\limits_{i = 1}^k {r_{U,i}^2}.
\end{aligned}
\label{eq58}\end{eqnarray}
The equality $(c)$ holds since the determinant of a triangular matrix is the product of the entries on the diagonal. Then, we have
\begin{align}
r_{U,k}^2 = \frac{{\prod\limits_{i = 1}^k {r_{U,i}^2} }}{{\prod\limits_{i = 1}^{k - 1} {r_{U,i}^2} }} = \frac{{\det ({\bf{V}}_{U,(k)}^T{{\bf{V}}_{U,(k)}})}}{{\det ({\bf{V}}_{U,(k - 1)}^T{{\bf{V}}_{U,(k - 1)}})}}.
\label{eq59}\end{align}
According to the Cauchy-Binet formula~\cite{Cauchy}, we can derive as
\begin{align}
\det ({\bf{V}}_{U,(K)}^T{{\bf{V}}_{U,(K)}}) = \sum\limits_{{\cal S}_K} {\det ({\bf{V}}_{U,[{\cal S}_K]}^T)\det ({{\bf{V}}_{U,[{\cal S}_K]}})},
\label{eq60}\end{align}
where ${{\cal S}_K}$ denotes the set of any $K$ combinations. ${\bf{V}}_{U,[{\cal S}_K]}$ is a $K$-by-$K$ sub-matrix of containing in set ${{\cal S}_K}$. It is easy to see that ${\bf{V}}_{U,[{\cal S}_K]}$ is also a Vandermonde matrix, and thus we have
\begin{align}
\det ({{\bf{V}}_{U,[{{\cal S}_K}]}}) = \prod\limits_{\{a < b\}\in{{\cal S}_K}} {({\eta _b} - {\eta _a})}.
\label{eq61}\end{align}
From the derivations in~\eqref{eq59},~\eqref{eq60}, and~\eqref{eq61}, the proof is completed.

\section{Proof of Lemma 1}
To prove the Lemma, we first introduce a property of the Fekete points in polynomial interpolation. Define the associated fundamental (or cardinal) Lagrange interpolating polynomial as
\begin{align}
{l_k}(x) = \prod\limits_{i \ne k} {\frac{{x - {\beta _i}}}{{\beta _k} - {\beta _i}}}.
\label{eq62}\end{align}
The polynomial function with degree of $K$ is given by
\begin{align}
g(x) = \sum\limits_{k \in {\cal K}} {{l_k}(x)g({\beta _k})}.
\label{eq63}\end{align}
Based on \eqref{eq63}, each entry of ${\bf{V}}_{U,(K)}$ is equivalent to
\begin{align}
{\eta _n^{i}} = \sum\limits_{k \in {\cal K}} {{l_k}({\eta _n}){\beta _k^{i}}}.
\label{eq64}\end{align}
In this way, we obtain the ${\bf{V}}_{U,(K)}$ as
\begin{align}
{{\bf{V}}_{U,(K)}} = {\bf{LB}},
\label{eq65}\end{align}  
where ${\bf{L}}=\{{l_k}({\eta _n})\}$, i.e., the entry in $n$th row and $k$th column of ${\bf{L}}$ is ${l_k}({\eta _n})$, and ${\bf{B}}$ is a Vandermonde matrix determined by ${\boldsymbol{\beta}} = ({\beta _1},{\beta _2},...,{\beta _K})$. Then, we rewrite the objective function in \textbf{P2.2} as
\begin{align} 
	{\mathscr{F}}_{{\cal S}_K}({\boldsymbol{\eta}})=\sum\limits_{{\cal S}_K} {\prod\limits_{\{a < b\}\in {{\cal S}_K}} {({\eta _b} - {\eta _a})}^2}  = \det ({\bf{V}}_{U,(K)}^T{\bf{V}}_{U,(K)}).
	\label{eq66}\end{align}
Inserting~\eqref{eq65} into~\eqref{eq66}, we arrive at
\begin{eqnarray}
	\begin{aligned}[b]
	&\det ({\bf{V}}_{U,(K)}^T{{\bf{V}}_{U,(K)}})\\&
	= \det \left( {{{({\bf{LB}})}^T}({\bf{LB}})} \right)\\&
	= \det ({{\bf{B}}^T}{{\bf{L}}^T}{\bf{LB}})\\&
    \overset{(d)} = \det ({\bf{B}}^T)\det ({{\bf{L}}^T}{\bf{L}})\det ({\bf{B}})\\&
	= {\mathscr{F}}_{{\cal S}_K}^2({\boldsymbol{\beta }})\det ({{\bf{L}}^T}{\bf{L}})\\&
	\overset{(e)}\le {\mathscr{F}}_{{\cal S}_K}^2({\boldsymbol{\beta }}){\prod\limits_{k \in {\cal K}} {\left( {\sum\limits_{n \in {\cal N}} {l_k^2({\eta _n})} } \right)}}\\&
	\overset{(f)}\le {{\mathscr{F}}_{{\cal S}_K}^2}({\boldsymbol{\beta }}){\left( {\frac{1}{K}\sum\limits_{k \in {\cal K}} {\sum\limits_{n \in {\cal N}} {l_k^2({\eta _n})} } } \right)^K}.
\end{aligned}
\label{eq67}\end{eqnarray}
The equality $(d)$ is obtained due to the invertible matrices ${\bf{L}}$ and ${\bf{B}}$, and the inequality $(e)$ and $(f)$ hold due to the Hadamard inequality~\cite{Hadamard} and arithmetic mean-geometric mean (AM-GM) inequality~\cite{inequality}, respectively. 

Considering the property of polynomial, we have
\begin{align}
\mathop {\max }\limits_{ - 1 \le x \le 1} \sum\limits_{k \in {\cal K}} {{l_k}(x)}  = 1.
\label{eq68}\end{align}
It can attain the maximum value for the Fekete-Gauss-Lobatto points~\cite[LM. 2.1.]{Fekete}. Consequently, substituting~\eqref{eq68} into~\eqref{eq67}, the upper bound can be obtained by
\begin{align}
{\mathscr{F}}_{{\cal S}_K}({\boldsymbol{\eta}}) \le {{\mathscr{F}}_{{\cal S}_K}^2}({\boldsymbol{\beta }}){\left( {\frac{N}{K}} \right)^K}.
\label{eq69}\end{align}
Lemma~\ref{lemma01} is thus proved.

\section{Proof of Lemma 2}
Based on the independent Eve channels, the left-side in~\eqref{eq11c} is equivalently recast as
\begin{eqnarray}
	&\Pr \left\{ {\mathop {\max }\limits_{q \in  {\cal Q}} {\sum\limits_{k \in  {\cal K}} {{\rm{Tr}}} \left( {{{\bf{H}}_{E,q}}{{\bf{W}}_k}} \right)}/{{\sigma _{E}^2}} \le {\xi }} \right\}\notag\\&
	= \prod\limits_{q \in  {\cal Q}} {\Pr \left\{ {\sum\limits_{k \in  {\cal K}} {{\rm{Tr}}} \left( {{{\bf{H}}_{E,q}}{{\bf{W}}_k}} \right)}/{{\sigma _{E}^2}} \le {\xi }\right\}},
	\label{eq70}\end{eqnarray}
where ${{\bf{H}}_{E,q}}={{\bf{h}}_{E,q}}{{\bf{h}}_{E,q}^H}$. Therefore, constraint~\eqref{eq11c} is equivalent to
\begin{align}
	&\Pr \left\{ {\mathop {\max }\limits_{q \in  {\cal Q}} {\sum\limits_{k \in  {\cal K}} {{\rm{Tr}}} \left( {{{\bf{H}}_{E,q}}{{\bf{W}}_k}} \right)}/{{\sigma _{E}^2}} \le {\xi }} \right\} \ge \kappa \label{eq71}\\&
	\Leftrightarrow \Pr \left\{ {{\rm{Tr}}\left( {{{\bf{H}}_{E}}{\bf{{Z}}}} \right) \le {\xi}\sigma _{E}^2} \right\} \ge {\kappa ^{1/Q}},
	\label{eq72} \end{align}
where ${\bf{{Z}}} = \sum\nolimits_{k \in {\cal K}} {{{\bf{W}}_k}}={\bf{W}}{{\bf{W}}^H}$. Considering the i.i.d. Eve channels, the index of Eve channels can be omitted. Then, we calculate the upper bound of the probabilistic constraint, satisfying
\begin{eqnarray}
	\begin{aligned}[b]
		{\mathop{\rm Tr}\nolimits} \left( {{{\bf{H}}_E}{\bf{Z}}} \right) & \overset{{(g)}}\le \sum\limits_{i \in {\cal N}} {{\lambda _i}} \left( {{{\bf{H}}_E}} \right){\lambda _i}({\bf{Z}})\\&
		\overset{{(h)}}= {\lambda _{\max }}\left( {{{\bf{H}}_E}} \right){\lambda _{\max }}({\bf{Z}})\\&
		\overset{{(i)}}= {\mathop{\rm Tr}\nolimits} \left( {{{\bf{H}}_E}} \right){\lambda _{\max }}({\bf{Z}}),
	\end{aligned}
	\label{eq73}\end{eqnarray}
where ${\lambda _{\max }}\left(  \cdot  \right)$ and ${\lambda _{i}}\left(  \cdot  \right)$ are the maximum and the $i$th largest eigenvalue of a matrix, respectively. The inequality $(g)$ utilizes the trace inequality of positive Hermitian matrices~\cite{Majorization}. The steps $(h)$ and $(i)$ exploit a rank-one positive semidefinite matrix. Combining~\eqref{eq72} and~\eqref{eq73}, we get
\begin{align}
	\Pr \left\{ {{\rm{Tr}}\left( {{{\bf{H}}_E}{\bf{Z}}} \right) \le {\xi}\sigma _E^2} \right\} \ge \Pr \left\{ {{\rm{Tr}}\left( {{{\bf{H}}_E}} \right){\lambda _{\max }}\left( {\bf{Z}} \right) \le {\xi}\sigma _E^2} \right\}.
	\label{eq74}\end{align}
Now let us return to constraint~\eqref{eq11c}, yielding
\begin{eqnarray}
	\begin{aligned}[b]
		&\Pr \left\{  {\mathop {\max }\limits_{q \in  {\cal Q}} {\sum\limits_{k \in  {\cal K}} {{\rm{Tr}}} \left( {{{\bf{H}}_{E,q}}{{\bf{W}}_k}} \right)}/{{\sigma _{E}^2}} \le {\xi }} \right\}\\&
		\ge \Pr \left\{ {{\rm{Tr}}\left( {{{\bf{H}}_E}} \right){\lambda _{\max }}\left( {\bf{Z}} \right) \le {\xi}\sigma _E^2} \right\} \ge {\kappa ^{1/Q}}\\&
		\overset{{(j)}} \Leftrightarrow \Pr \left\{ {\frac{{{\lambda _{\max }}\left( {\bf{Z}} \right)}}{{{\xi}\sigma _E^2}} \le \frac{1}{{{\rm{Tr}}\left( {{{\bf{H}}_E}} \right)}}} \right\} \ge {\kappa ^{1/Q}}\\&
		\overset{{(k)}} \Leftrightarrow \Pr \left\{ {\frac{1}{{{\rm{Tr}}\left( {{{\bf{H}}_E}} \right)}} \le \frac{{{\lambda _{\max }}\left( {\bf{Z}} \right)}}{{{\xi}\sigma _E^2}}} \right\} \le 1 - {\kappa ^{1/Q}}\\&
		\overset{{(l)}} \Leftrightarrow {\lambda _{\max }}\left( {\bf{Z}} \right) \le \Phi _N^{ - 1}\left( {1 - {\kappa ^{1/Q}}} \right){\xi}\sigma _E^2\\&
		\Leftrightarrow {\bf{Z}} \preceq {{\bf{I}}_N}\left[ {\Phi _N^{ - 1}\left( {1 - {\kappa ^{1/Q}}} \right){\xi}\sigma _E^2} \right].
	\end{aligned}
	\label{eq75}\end{eqnarray}
The equivalent transform $(j)$ and $(k)$ utilize the positive definiteness of matrix ${{\bf{H}}_E}$ and the basic property of probability, respectively. Transform $(l)$ is performed similarly to the derivations in~\cite[LM. 1]{Cognitive}. $\Phi _N^{ - 1}\left( \cdot \right)$ indicates the inverse c.d.f. of an inverse central chi-square random variable. Therefore, the proof of Lemma~\ref{lemma02} is concluded.

\ifCLASSOPTIONcaptionsoff
  \newpage
\fi

\bibliographystyle{IEEEtran}
\bibliography{IEEEabrv,REF}

\end{document}